\documentclass[sigconf]{acmart}

\usepackage[normalem]{ulem}
\usepackage{array}
\usepackage{graphicx}
\usepackage{textcomp}
\usepackage{ifthen}
\usepackage{xcolor}
\usepackage{blindtext}
\usepackage{listings}
\usepackage{xspace}
\usepackage{balance}

\usepackage{footnote}
\usepackage{comment}
\usepackage{enumitem,kantlipsum}
\makesavenoteenv{tabular}
\makesavenoteenv{table}
\usepackage{booktabs} 
\usepackage{multicol}
\usepackage{multirow}
\usepackage{wrapfig}
\usepackage{hyperref}
\usepackage{svg}
\usepackage{caption}
\usepackage{mathtools}
\usepackage[symbol]{footmisc}
\usepackage{cleveref}
\usepackage{subfig}
\usepackage{microtype}
\usepackage{amsmath,amsfonts}
\usepackage{algorithmic}

\newcommand{\company}{Meta\xspace}
\newcommand{\sys}{Neo\xspace}

\newcommand{\alltoall}{\texttt{AlltoAll}\xspace}
\newcommand{\allreduce}{\texttt{AllReduce}\xspace}
\newcommand{\reducescatter}{\texttt{ReduceScatter}\xspace}
\newcommand{\onetomany}{\texttt{OneToMany}\xspace}
\newcommand{\manytomany}{\texttt{ManyToMany}\xspace}
\newcommand{\zion}{\textit{Zion}\xspace}
\newcommand{\zionex}{\textit{ZionEX}\xspace}

\newcommand{\zj}[1]{\textcolor{red}{(Zhihao: #1)}}

\copyrightyear{2022} 
\acmYear{2022} 
\setcopyright{acmlicensed}\acmConference[ISCA '22]{The 49th Annual International Symposium on Computer Architecture}{June 18--22, 2022}{New York, NY, USA}
\acmBooktitle{The 49th Annual International Symposium on Computer Architecture (ISCA '22), June 18--22, 2022, New York, NY, USA}
\acmPrice{15.00}
\acmDOI{10.1145/3470496.3533727}
\acmISBN{978-1-4503-8610-4/22/06}

\renewcommand{\shortauthors}{D. Mudigere, Y. Hao, J. Huang, and Z. Jia et al.}

\begin{document}

\title{Software-Hardware Co-design for Fast and Scalable Training of Deep Learning Recommendation Models}
\subtitlenote{This paper is part of the Industry Track of ISCA 2022's program.}

\author{\large Dheevatsa Mudigere\textsuperscript{$\dagger\ddagger$}\authornote{These authors contributed equally.}, Yuchen Hao\textsuperscript{$\dagger\ddagger$}\authornotemark[1], Jianyu Huang\textsuperscript{$\dagger\ddagger$}, Zhihao Jia\textsuperscript{$\S$}, Andrew Tulloch\textsuperscript{$\ddagger$}, \\ Srinivas Sridharan\textsuperscript{$\ddagger$}, Xing Liu\textsuperscript{$\ddagger$}, Mustafa Ozdal\textsuperscript{$\ddagger$}, Jade Nie\textsuperscript{$\ddagger$}, Jongsoo Park\textsuperscript{$\ddagger$}, Liang Luo\textsuperscript{$\ddagger$}, Jie (Amy) Yang\textsuperscript{$\ddagger$}, Leon Gao\textsuperscript{$\ddagger$}, Dmytro Ivchenko\textsuperscript{$\ddagger$}, Aarti Basant\textsuperscript{$\ddagger$}, Yuxi Hu\textsuperscript{$\ddagger$}, Jiyan Yang\textsuperscript{$\ddagger$}, Ehsan K. Ardestani\textsuperscript{$\ddagger$}, Xiaodong Wang\textsuperscript{$\ddagger$}, Rakesh Komuravelli\textsuperscript{$\ddagger$}, Ching-Hsiang Chu\textsuperscript{$\ddagger$}, Serhat Yilmaz\textsuperscript{$\ddagger$}, Huayu Li\textsuperscript{$\ddagger$}, Jiyuan Qian\textsuperscript{$\ddagger$}, Zhuobo Feng\textsuperscript{$\ddagger$}, Yinbin Ma\textsuperscript{$\ddagger$}, Junjie Yang\textsuperscript{$\ddagger$}, Ellie Wen\textsuperscript{$\ddagger$}, Hong Li\textsuperscript{$\ddagger$}, Lin Yang\textsuperscript{$\ddagger$}, Chonglin Sun\textsuperscript{$\ddagger$}, Whitney Zhao\textsuperscript{$\ddagger$}, Dimitry Melts\textsuperscript{$\ddagger$}, Krishna Dhulipala\textsuperscript{$\ddagger$}, KR Kishore\textsuperscript{$\ddagger$}, Tyler Graf\textsuperscript{$\ddagger$}, Assaf Eisenman\textsuperscript{$\ddagger$}, Kiran Kumar Matam\textsuperscript{$\ddagger$}, Adi Gangidi\textsuperscript{$\ddagger$}, Guoqiang Jerry Chen\textsuperscript{$\ddagger$}, Manoj Krishnan\textsuperscript{$\ddagger$}, Avinash Nayak\textsuperscript{$\ddagger$}, Krishnakumar Nair\textsuperscript{$\ddagger$}, Bharath Muthiah\textsuperscript{$\ddagger$}, Mahmoud khorashadi\textsuperscript{$\ddagger$}, Pallab Bhattacharya\textsuperscript{$\ddagger$}, Petr Lapukhov\textsuperscript{$\ddagger$}, Maxim Naumov\textsuperscript{$\ddagger$}, Ajit Mathews\textsuperscript{$\ddagger$}, Lin Qiao\textsuperscript{$\ddagger$}, Mikhail Smelyanskiy\textsuperscript{$\ddagger$}, Bill Jia\textsuperscript{$\ddagger$}, Vijay Rao\textsuperscript{$\ddagger$} \vspace{4pt}}
\affiliation{%
  \institution{\vspace{0.3em}
    {\Large \textsuperscript{$\ddagger$}Meta Platforms, \textsuperscript{$\S$}Carnegie Mellon University}
  \vspace{0.5em}}
 \country{}
}

\renewcommand{\authors}{Dheevatsa Mudigere, Yuchen Hao, Jianyu Huang, Zhihao Jia, Andrew Tulloch, Srinivas Sridharan, Xing Liu, Mustafa Ozdal, Jade Nie, Jongsoo Park, Liang Luo, Jie (Amy) Yang, Leon Gao, Dmytro Ivchenko, Aarti Basant, Yuxi Hu, Jiyan Yang, Ehsan K. Ardestani, Xiaodong Wang, Rakesh Komuravelli, Ching-Hsiang Chu, Serhat Yilmaz, Huayu Li, Jiyuan Qian, Zhuobo Feng, Yinbin Ma, Junjie Yang, Ellie Wen, Hong Li, Lin Yang, Chonglin Sun, Whitney Zhao, Dimitry Melts, Krishna Dhulipala, KR Kishore, Tyler Graf, Assaf Eisenman, Kiran Kumar Matam, Adi Gangidi, Guoqiang Jerry Chen, Manoj Krishnan, Avinash Nayak, Krishnakumar Nair, Bharath Muthiah, Mahmoud khorashadi, Pallab Bhattacharya, Petr Lapukhov, Maxim Naumov, Ajit Mathews, Lin Qiao, Mikhail Smelyanskiy, Bill Jia, Vijay Rao}

\renewcommand{\shortauthors}{D. Mudigere, Y. Hao, J. Huang, and Z. Jia et al.}




\begin{abstract}
Deep learning recommendation models (DLRMs) have been used across many business-critical services at \company and are the single largest AI application in terms of infrastructure demand in its data-centers.
In this paper, we present \sys, a software-hardware co-designed system for high-performance distributed training of large-scale DLRMs.
\sys employs a novel 4D parallelism strategy that combines table-wise, row-wise, column-wise, and data parallelism for training massive embedding operators in DLRMs.
In addition, \sys enables extremely high-performance and memory-efficient embedding computations using a variety of critical systems optimizations, including hybrid kernel fusion, software-managed caching, and quality-preserving compression.
Finally, \sys is paired with \zionex, a new hardware platform co-designed with \sys's 4D parallelism for optimizing communications for large-scale DLRM training.
Our evaluation on 128 GPUs using 16 \zionex nodes shows that \sys outperforms existing systems by up to 40$\times$ for training 12-trillion-parameter DLRM models deployed in production.
\end{abstract}

\maketitle

\section{Introduction}
Deep learning recommendation models (DLRMs) are ubiquitously used by online companies, including Amazon for selecting items in its catalog \cite{amazon_reco, yifei2020, lopez2021bandits}, Netflix for showing movie options \cite{netflix_reco, koren2009matrix}, and Google for displaying personalized advertisements \cite{goog_widedeep, covington2016deep, he2017neural}.

They have also been adopted by standard benchmarking organizations, such as MLCommons (MLPerf)~\cite{mlperfTraining, mlperfInference}.
At \company, we have been using recommendation models extensively for ranking and click through rate (CTR) prediction, including news feed and search services~\cite{hazelwood2018applied, naumov2020deep, park2018deep, udit9065589}. DLRMs are the single largest AI application in terms of infrastructure demand in data centers.

\begin{figure}
     \centering
     \subfloat{
         \centering
         \includegraphics[width=210pt]{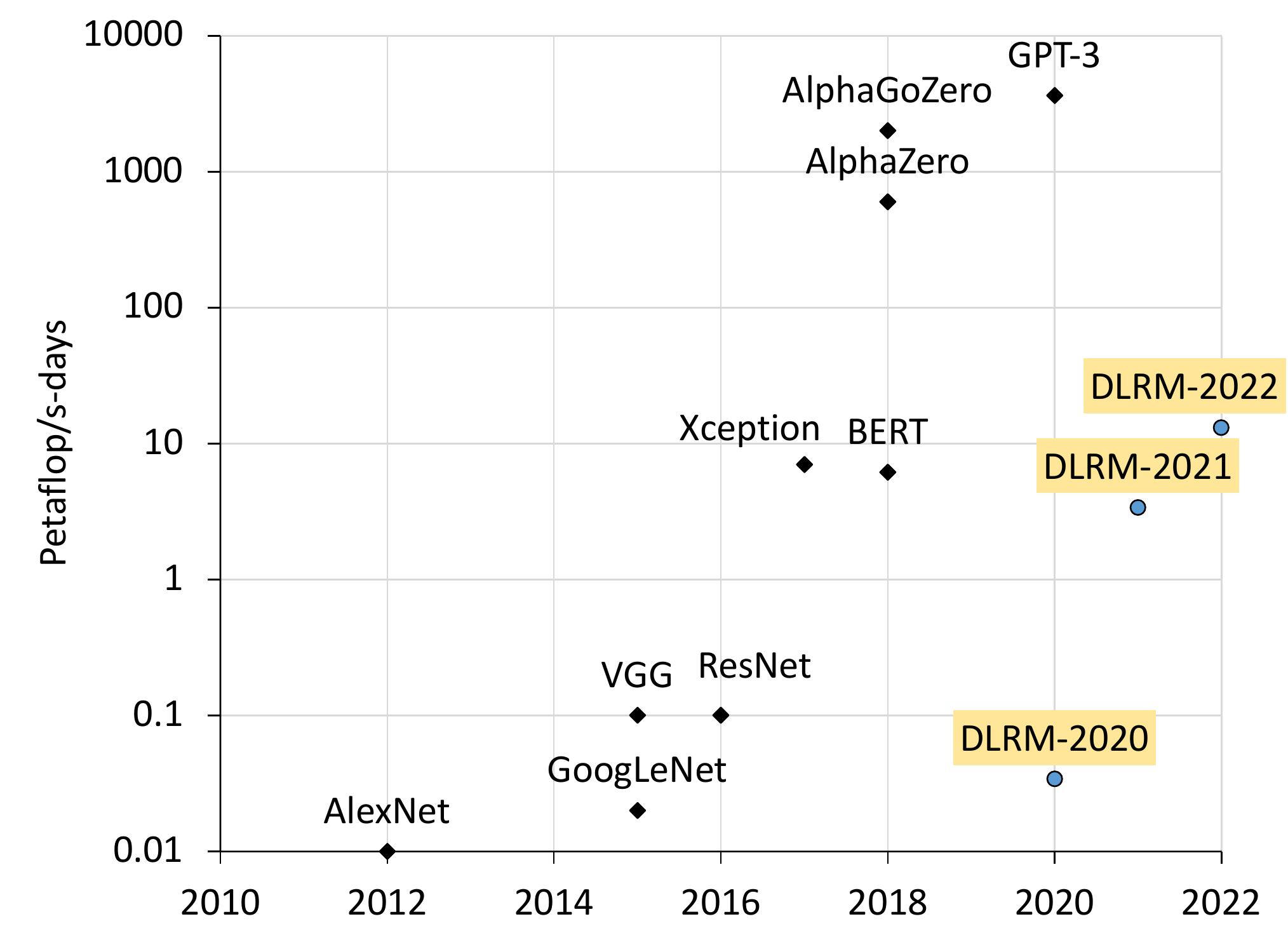}
         \label{fig:comp_totalCompute}
     }\\
    \subfloat{
         \centering
         \includegraphics[width=210pt]{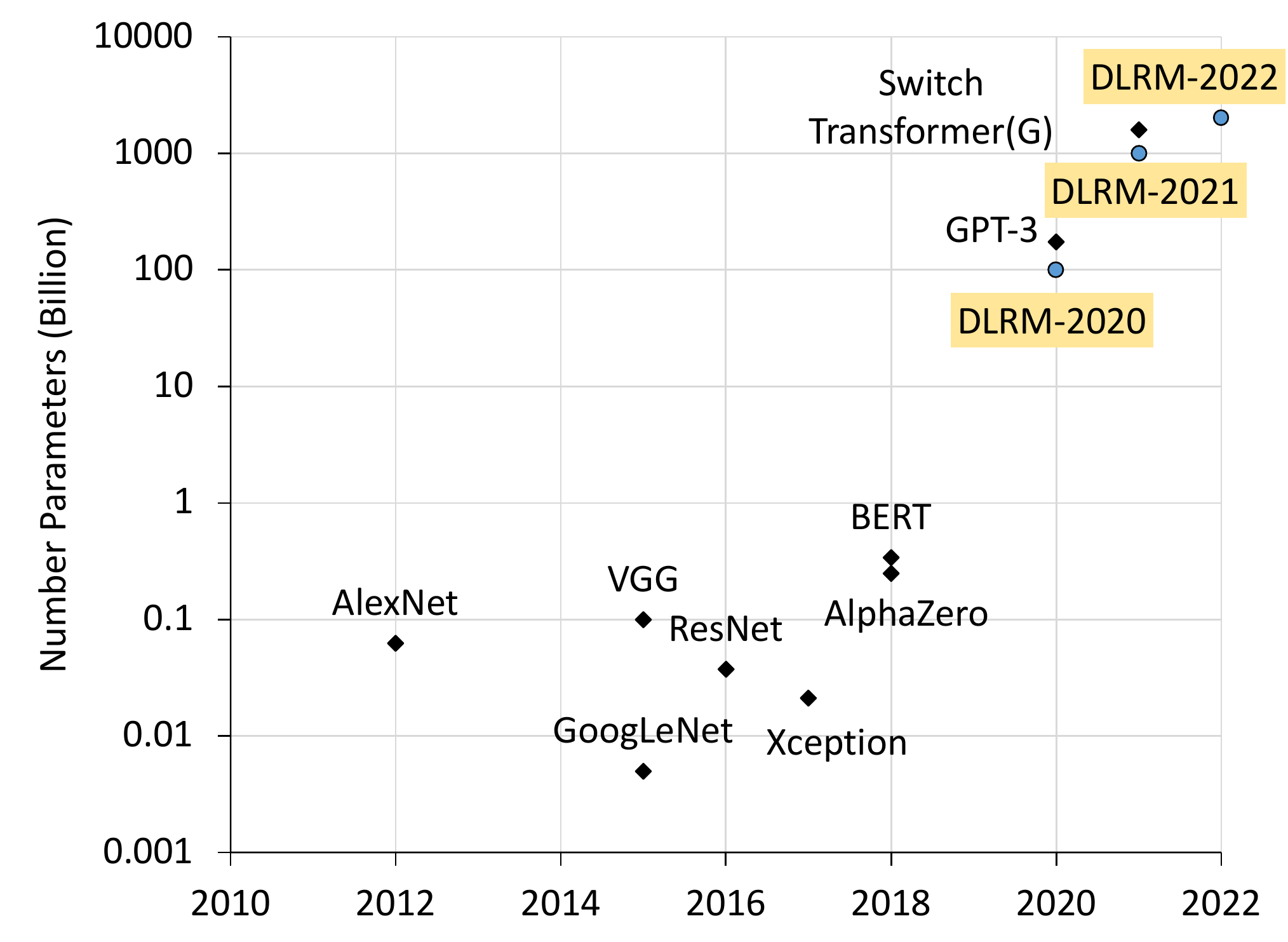}
         \label{fig:comp_params}
    }
    \caption{Comparing deep learning models in total amount of compute, in petaflop/s-days (top)~\cite{ai-and-compute} and model capacity (bottom).}
    \label{fig:comp_graphs}
\end{figure}

Unlike conventional deep neural networks (DNNs) with mainly compute-intensive operators (e.g., convolution and matrix multiplication), DLRMs combine compute-intensive components with up to thousands of data-intensive {\em embedding operators}, each with a different resource requirement and performance characteristic~\cite{DLRM19}.
As a result, DLRMs generally exhibit much lower arithmetic intensity and larger model sizes compared to their computer vision \cite{resnet,googlenet,xception}, natural language processing \cite{transformer,gpt3,bert}, and reinforcement learning counterparts \cite{alphago,alphazero}, with models having trillions of parameters being deployed in practice, as shown in~\Cref{fig:comp_graphs}.

Existing software and hardware solutions tailored for DNNs achieve only suboptimal performance and limited scalability on DLRMs due to the following software/hardware limitations.


On the software side, existing deep learning frameworks parallelize DNN training typically using either {\em data}, {\em model} or {\em pipeline parallelism}~\cite{tensorflow, pytorch, pytorchddp}. 
Frameworks that support combinations of these strategies are generally designed for specific DNN applications~\cite{deepspeed, flexflow, pipedream, megatron}.
However, existing parallelization strategies designed and optimized for compute-intensive DNN models achieve limited performance and scalability for DLRMs.
In particular, data parallelism requires each device to save a replica of the entire model and therefore does not support DLRMs with up to trillions of parameters~\cite{pytorchddp}.
Moreover, a DLRM cannot be directly parallelized using model or pipeline parallelism due to the data-dependent behavior of its embedding operators.
Specifically, processing different training samples may require accesses to different embedding parameters depending on the categorical inputs of each sample.
This data-dependent behavior makes it infeasible to statically partition a DLRM's trainable parameters into disjoint subsets while satisfying data dependencies for all samples, a necessity for using model and pipeline parallelism.

In addition, today's DNN frameworks are designed and optimized for compute-intensive DNN computations and miss critical optimizations for data-intensive embedding operators.
Specifically, DLRMs contain up to thousands of embedding operators. The forward processing, backward propagation, and gradient synchronization for these embedding operators require launching thousands of CUDA kernels in a training iteration and consume up to terabytes of aggregated GPU device memory, introducing significant runtime overheads and memory requirements.

On the hardware side, modern hardware platforms such as GPU-based clusters provide significant capability boost, but they are not designed to match the performance characteristics of DLRMs.
Specifically, hardware platforms for DNN training are generally optimized for centralized inter-node communications (e.g., parameter servers~\cite{tensorflow}) and/or \allreduce communications (e.g., Horovod~\cite{horovod} and NCCL~\cite{nccl}).
However, as identified in~\Cref{sec:overview}, performant and scalable DLRM training requires efficient hardware support for a mixture of diverse communication patterns, including \allreduce, \alltoall, \reducescatter, \onetomany, and \manytomany.

\subsection{Our Approach}
We present \sys, a software-hardware co-designed system for fast and scalable DLRM training building on top of three key techniques.

\paragraph{4D parallelism.}
To enable fast and scalable training of the massive embedding operators in DLRMs, it is crucial to effectively balance the workload distribution across GPUs while minimizing communication costs.
We introduce a {\em 4D parallelism} strategy that combines {\em table-wise}, {\em row-wise}, {\em column-wise}, and {\em data} parallelism to jointly optimize the parallelization performance of embedding operators.
Additionally, \sys also supports applying 4D parallelism in a {\em recursive} manner at different levels of hardware hierarchy to further improve load balance and hardware efficiency.

\paragraph{High-performance embedding computation.}
\sys employs two novel optimizations to minimize the computational costs and memory requirements of embedding operators.
First, we introduce a {\em hybrid kernel fusion} technique that fuses (1) multiple embedding operators and (2) embedding computations and their parameter updates {\em all} in a single CUDA kernel.
This is realized by co-designing the optimization algorithms and software implementation of embedding operators.
Second, to provide sufficient memory capacity for DLRM training, \sys uses a {\em software-managed caching} mechanism to leverage the memory hierarchy of modern hardware platforms.
Finally, a variety of compression techniques~\cite{lowprecision,koren2009matrix} are further applied to minimize memory requirements.

\paragraph{Hardware platform design.}
We introduce \zionex, a new hardware platform co-designed with \sys's 4D parallelism to optimize inter-node communications for distributed DLRM training.
\zionex supports a {\em fully-connected} topology across all GPUs in the cluster by using a dedicated \emph{RDMA over Converged Ethernet} (RoCE) based scale-out network.
This topology design promotes high-performance data transfers for the performance-dominating communication workloads (e.g., \alltoall and \manytomany) in distributed DLRM training.
Meanwhile, \zionex supports both the RDMA and GPUDirect communication protocols and retains flexible intra-node GPU fabric.
This enables high-performance DLRM training on \zionex, while ensuring compatibility with existing data-center infrastructure to allow wide deployment of \zionex.

\paragraph{Results.} We have evaluated \sys on three DLRMs deployed in production for different tasks, including click through rate prediction, ranking, and engagement, representing a diverse set of production-level recommendation models.
Our evaluation on 128 A100 GPUs on 16 \zionex nodes shows that \sys is able to process up to 1.7 million queries per second for training DLRMs with 12 trillion parameters, a 40$\times$ speedup compared to existing solutions for DLRM training in production.
Ablation studies show that 4D parallelism, high-performance embedding computation, and the new \zionex platform are all critical to enabling fast and scalable DLRM training. 
\\ \newpage
To summarize, our contributions are:
\begin{itemize}
    \item We present \sys, a software-hardware co-designed system for fast and scalable training of DLRMs. \sys outperforms existing systems by up to 40$\times$ for training large-scale DLRMs with 12 trillion parameters.
    \item We propose 4D parallelism, a combination of
    table-wise, row-wise, column-wise, and data parallelism for training embedding operators.
    \item We develop and implement high-performance embedding operators using hybrid kernel fusion, software-managed caching, and quality-preserving compression.
    \item We build \zionex, a new hardware platform co-designed with \sys's 4D parallelism to accelerate a variety of communication patterns in DLRM training.
\end{itemize}

\section{Background}
\label{sec:background}

\begin{table}
\centering
\small
\caption{\small{Sample DLRM time to train latency resources demand}}
\label{tab:resource}
\begin{tabular}{l|c}
\toprule
Total compute & $\phantom{1.}$1+ PF/s  \\ \midrule
Total memory capacity & $\phantom{.}$1+ TB \\ \midrule
Total memory BW & 100+ TB/s \\ \midrule
Network injection BW per worker & 100+ GB/s \\ \midrule
Network bisection BW & $\phantom{1.}$1+ TB/s \\ \midrule
\end{tabular}
\end{table}

DLRMs typically have two modes of training - offline and online, each with varying requirements.
The offline training can be viewed more as a pre-training, where a candidate model is trained on sufficiently large historical data, and expected to generalize when deployed to current/unseen samples. Once deployed, DLRMs continue to be trained in an online mode using the data it has already served on. Offline training is throughput limited, fitting into the more conventional \emph{"train as fast as possible on as much data as possible"} paradigm, whereas online training is more latency sensitive, with the frequency of re-training and update being an important factor. For online training, the throughput requirement is lower hence it might be desired to use proportionally lower resources. 
This creates an unique requirement of training very large models at smaller scales capable of tolerating lower throughput.


This paper focuses on offline training with more demanding training throughput needs --- up to millions of samples (queries) per second resulting from processing through tens of petabytes of training data within a reasonable time. This drives the training platform requirements, as summarized in Table~\ref{tab:resource}. 

\begin{figure}
    \centering
    \includegraphics[scale=0.33]{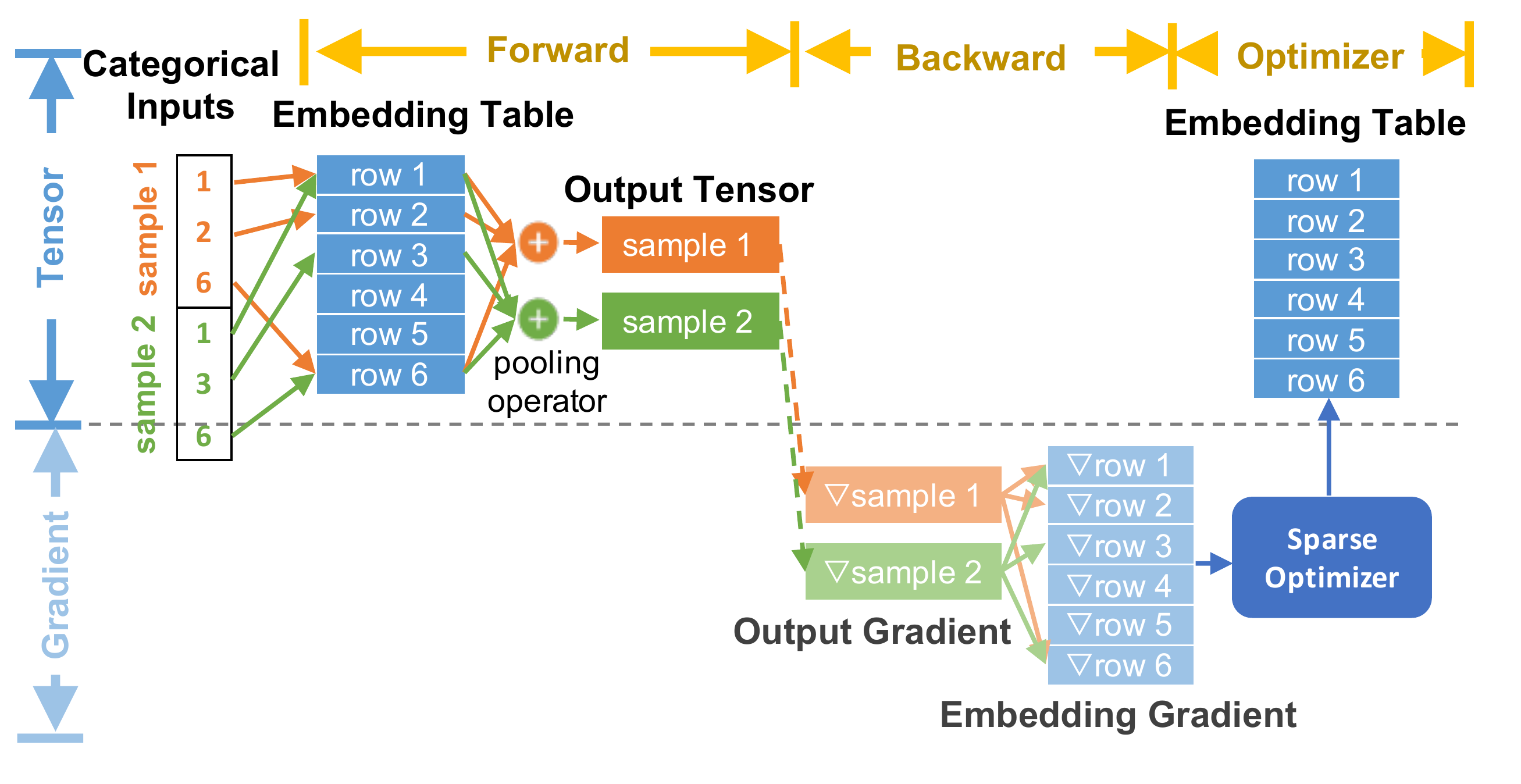}
    \caption{Workflow of an embedding operator.}
    \label{fig:embedding}
\end{figure}

\paragraph{Embedding operators.}
A major difference between DLRMs and conventional deep neural networks is leveraging categorical features such as users, posts, or pages.
The DLRMs used in production typically contain up to thousands of categorical features, each of which corresponds to a dedicated {\em embedding operator}.
An embedding operator takes as an input a multi-hot vector, and each non-zero element in the vector triggers a full row retrieval in the embedding table where each index in the input vector corresponds to a table row. Finally, all embedding rows for a given input vector are combined with element-wise pooling, as shown in Fig.~\ref{fig:embedding}.

\begin{figure}
    \centering
    \includegraphics[width=\linewidth]{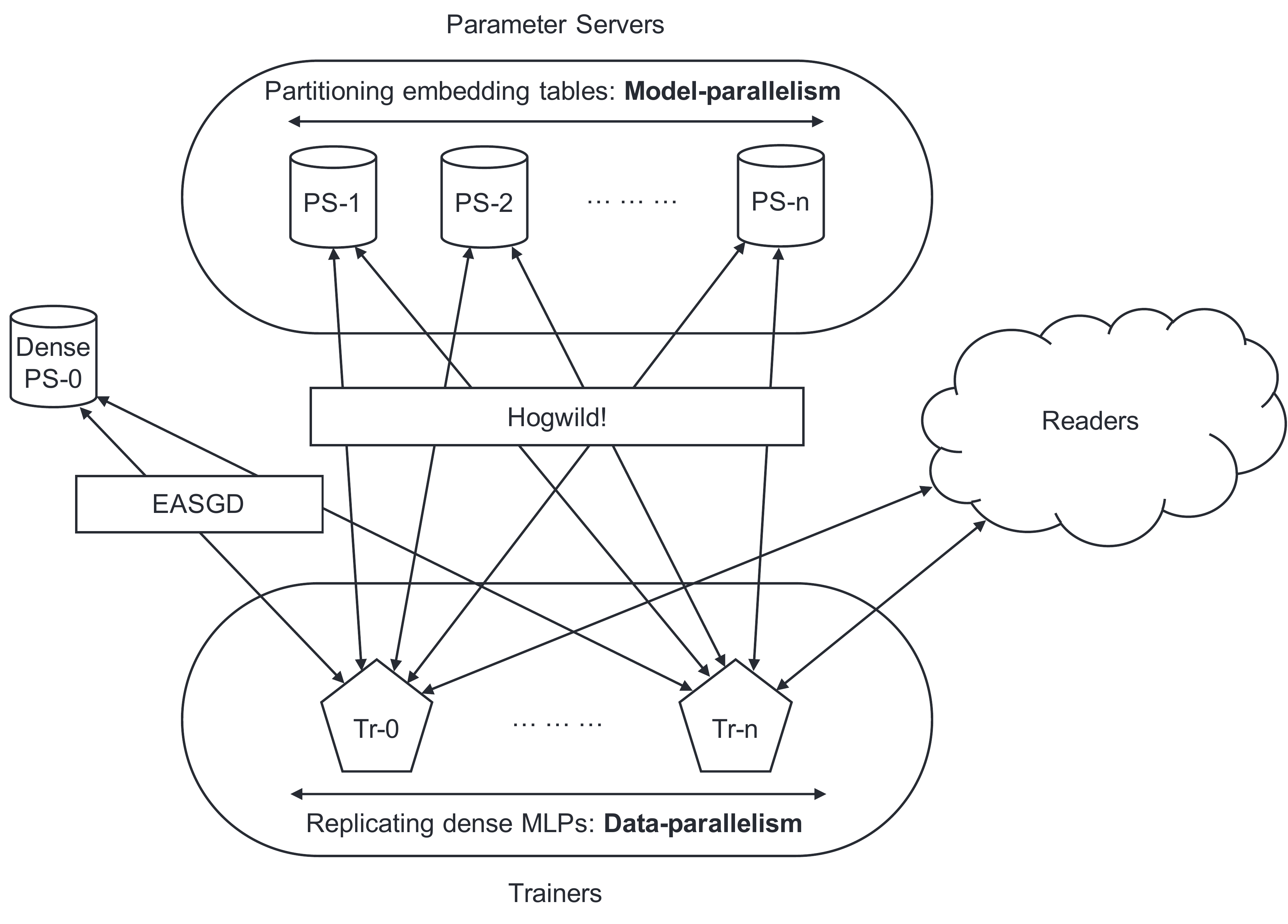}
    \caption{Disaggregated parameter-server based system}
    \label{fig:distributed-cpu}
\end{figure}

\paragraph{Parallelization strategies.} Traditionally a disaggregated parameter-server (PS) based distributed CPU training system has been used for training DLRMs in a production setting~\cite{hazelwood2018applied, naumov2020deep}. Specifically, the dense parameters from the MLP modules are duplicated between the trainers to exploit data-parallelism. Their weights are synchronized with a centralized dense parameter server using Elastic Averaging method SGD~\cite{easgd, Zheng2020}. On the other hand, The parameters from the embedding tables are partitioned and placed on multiple PS to exploit model-parallelism, since the size of embedding parameters simply prevents model replication. To maximize training throughput, the parameters of embedding operators are updated using Hogwild!~\cite{hogwild}. In addition, the readers are deployed on a separate tier of machines to feed training batches to the trainers as illustrated in Fig.~\ref{fig:distributed-cpu}.

Such PS-based system is well suited for DLRMs allowing scaling different components separately and achieving a balanced resource utilization when training different models with different trainer, parameter server and reader configurations. Moreover, resources in the system are largely fungible, making it low-cost for datacenter operations. 

However, the need for supporting DLRMs with trillions of parameters and therefore terabytes in size poses a serious challenge to the scalability of this approach, necessitating a steep increase of the number of trainers and parameter-servers to meet the ever growing training requirements.
This quickly becomes intractable, degrading model accuracy with staleness due to increased asynchronous updates across a very large number of workers. To tackle these issues, we build a high-performance synchronous training solution for large DLRMs, decoupling distributed scaling from statistical quality.

The efficient design of the synchronous training system leads us to use a novel combination of {\em 4D parallelism} (\Cref{sec:sharding}) for memory intensive embeddings tables, data parallelism for compute intensive DNN operators, and pipelining across different components. This hybrid parallelism requires \alltoall communications for the embedding lookup results~\cite{DLRM19, naumov2020deep}, as well as embedding table input redistribution if the inputs are streamed from database in batches, which is often the case.
Unlike \allreduce communications for gradient synchronizations, which can be overlapped, these \alltoall communications are on the critical path due to data dependencies, stressing the performance of the interconnect and communication primitives.  Furthermore DLRMs are typically trained on \emph{very large} amounts of data, which corresponds to mostly unstructured and unlabeled interactions from a wide variety of applications. Typical data-set sizes are in the range of several petabytes, necessitating the use of common, distributed network storage, such as the Tectonic filesystem \cite{fbwarmstorage}. For training, this data would need to be streamed in, putting additional stress on the host network and host-to-device bandwidth.
\section{Overview}
\label{sec:overview}
\begin{figure}
    \centering
    \includegraphics[width=\linewidth]{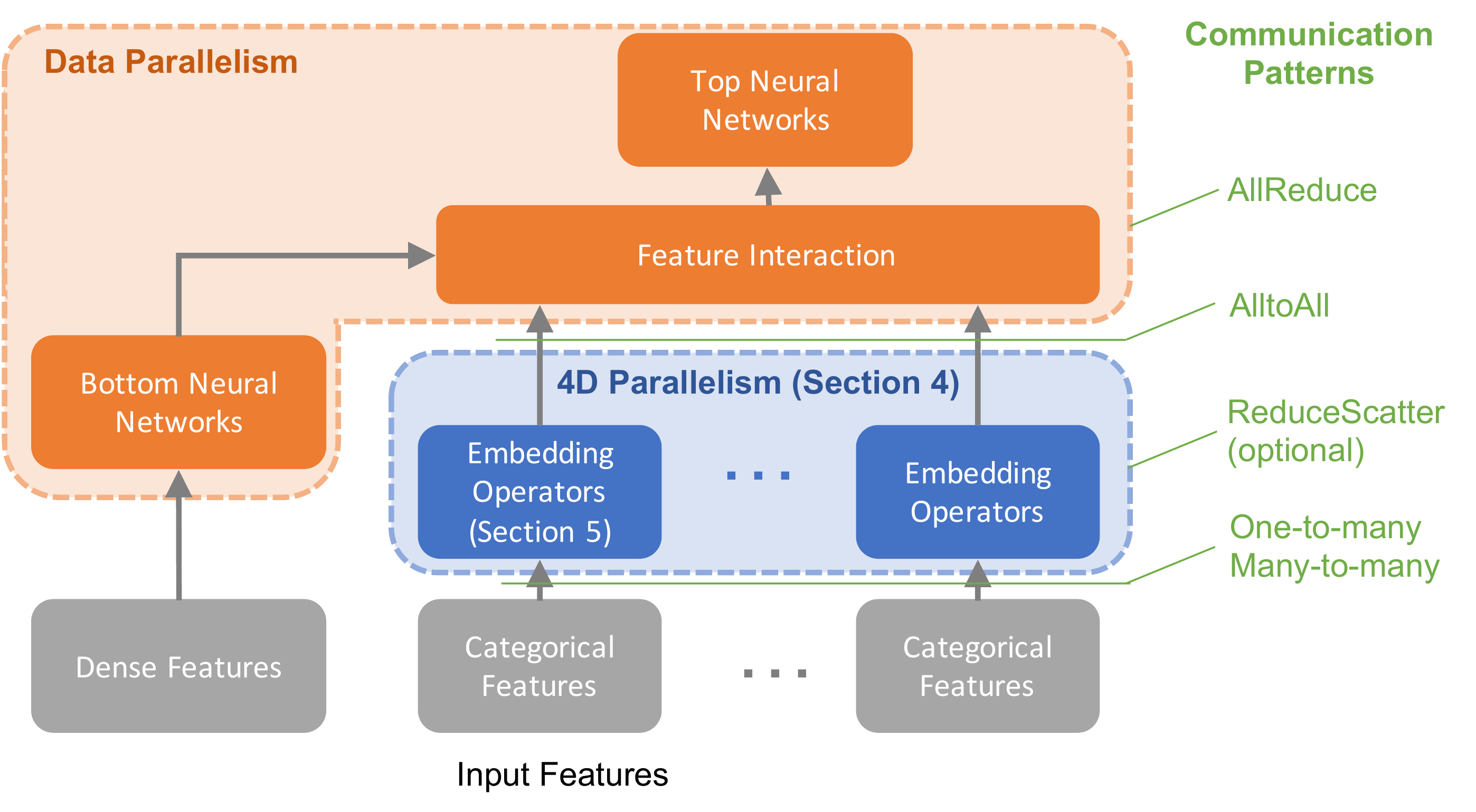}
    \caption{Neo overview. Each box in the figure indicates a neural network component, while edges between boxes are tensors shared between different components.}
    \label{fig:hpc_training}
\end{figure}

Fig.~\ref{fig:hpc_training} shows an overview of \sys, a software-hardware co-designed system for fast and scalable training of DLRMs.
This section briefly describes the key components of \sys.

First, \sys uses {\em data parallelism} for training compute-intensive DNN layers (shown in orange) and switches to a {\em 4D parallelism} strategy that combines {\em table-wise}, {\em row-wise}, {\em column-wise}, and {\em data} parallelism for efficient training of memory-intensive embedding operators.

Second, \sys is equipped with a high-performance implementation for embedding operators.
This is achieved by a number of critical systems optimizations, including (1) a {\em hybrid kernel fusion} technique to reduce the computational cost of embedding operators, (2) a {\em software-managed caching} mechanism to leverage heterogeneous memories of modern hardware platforms, and (3) a variety of {\em quality-preserving compression} techniques to minimize the memory requirement for embedding computation. 


Finally, \sys is deployed on \zionex, a new hardware platform co-designed with \sys's 4D parallelism to optimize inter-node communications for DLRM training.

Additionally, data I/O is an integral part of any training system, especially with the adoption of fully synchronous training and accelerators. First, the host to device transfer should be non-blocking and fast enough not to limit the overall training throughput. Ideally overlapping the input data transfers with training using double buffering or pipelining. Second, even though mapping input data distribution to collective communications between trainers is faster, this introduces additional challenges for the input and output data layout of the collective communications. Initial experiments show that these could add significant latency to the critical path. We will illustrate how we overcome these practical challenges in~\Cref{sec:data}.
\section{4D Parallelism} \label{sec:sharding}
\begin{figure*}
\centering
\subfloat[Table-wise Parallelism] {
\includegraphics[scale=0.43]{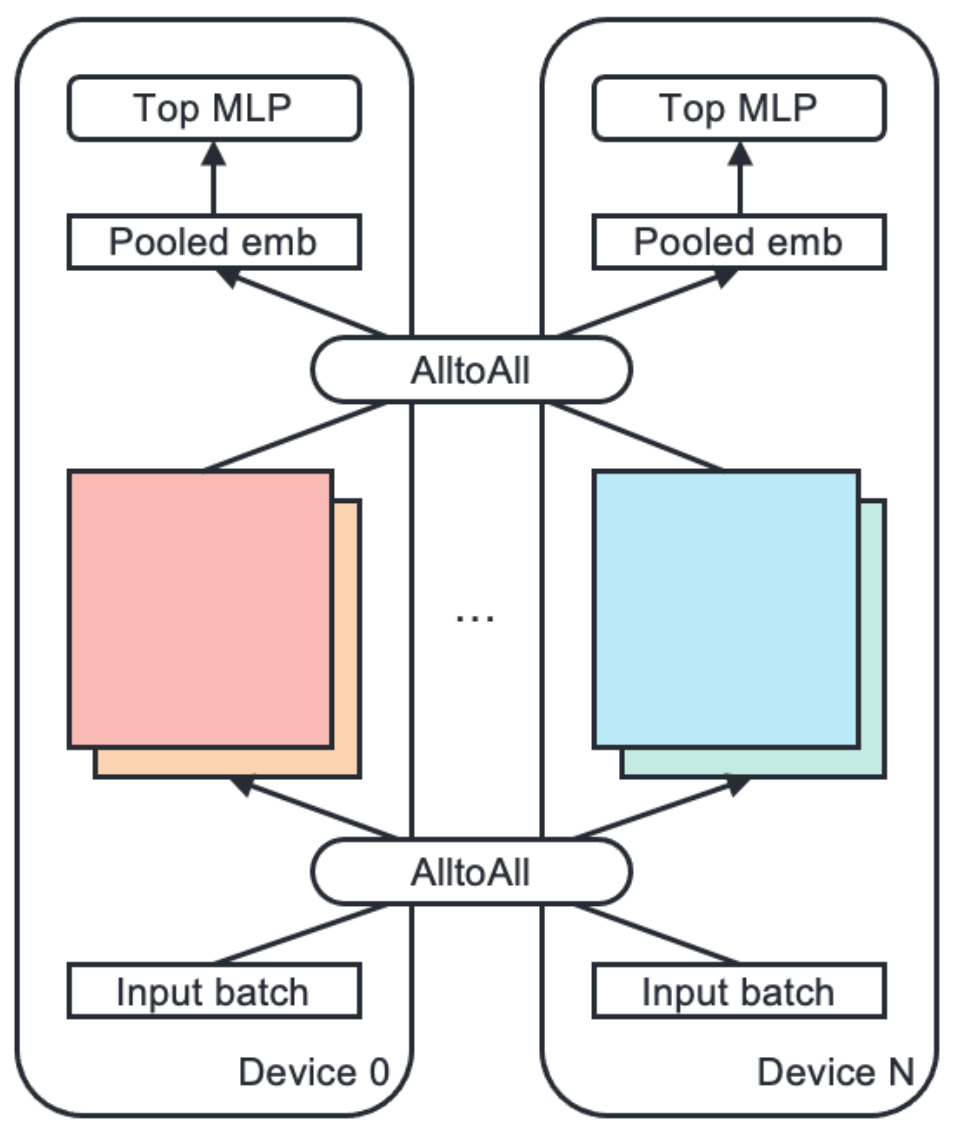}
\label{fig:sharding_tw}
}
\subfloat[Row-wise Parallelism] {
\includegraphics[scale=0.43]{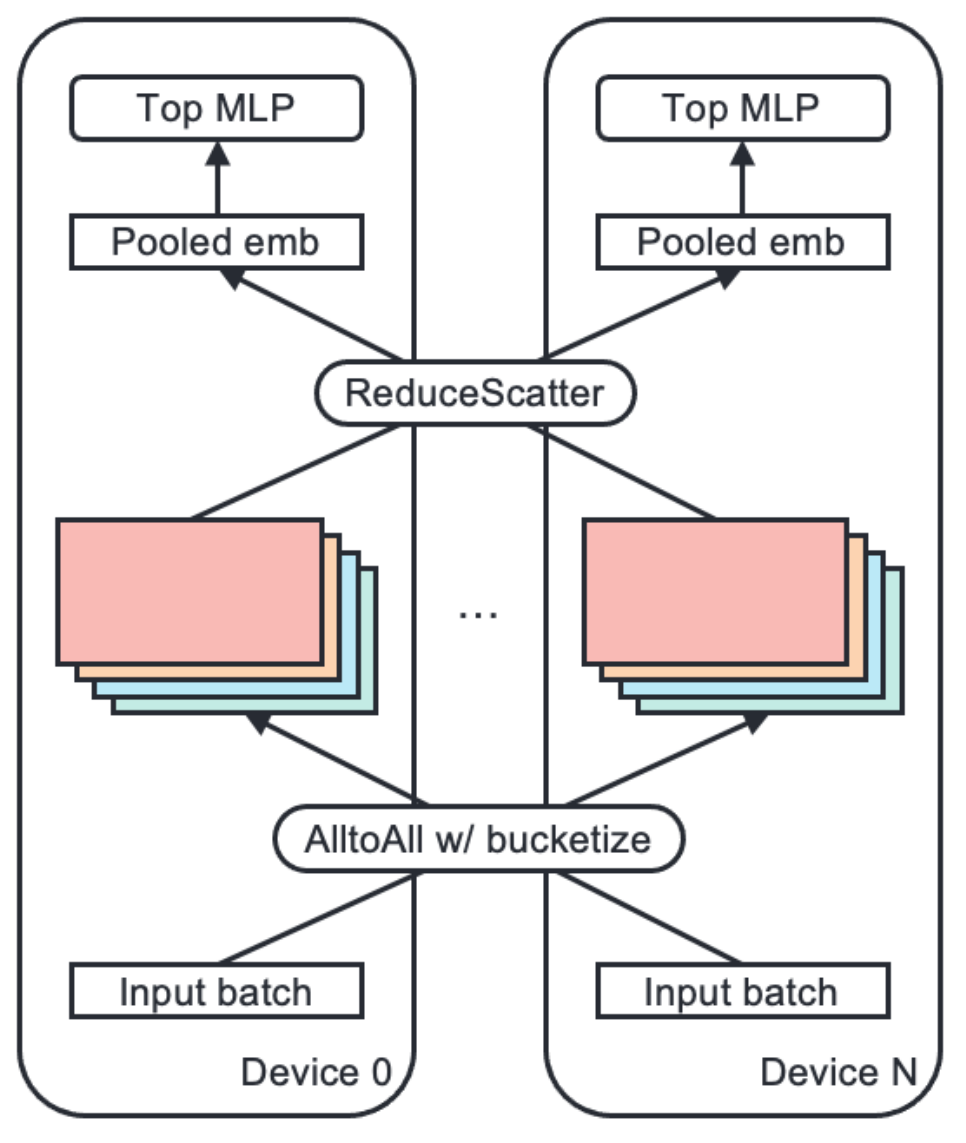}
\label{fig:sharding_rw}
}
\subfloat[Column-wise Parallelism] {
\includegraphics[scale=0.43]{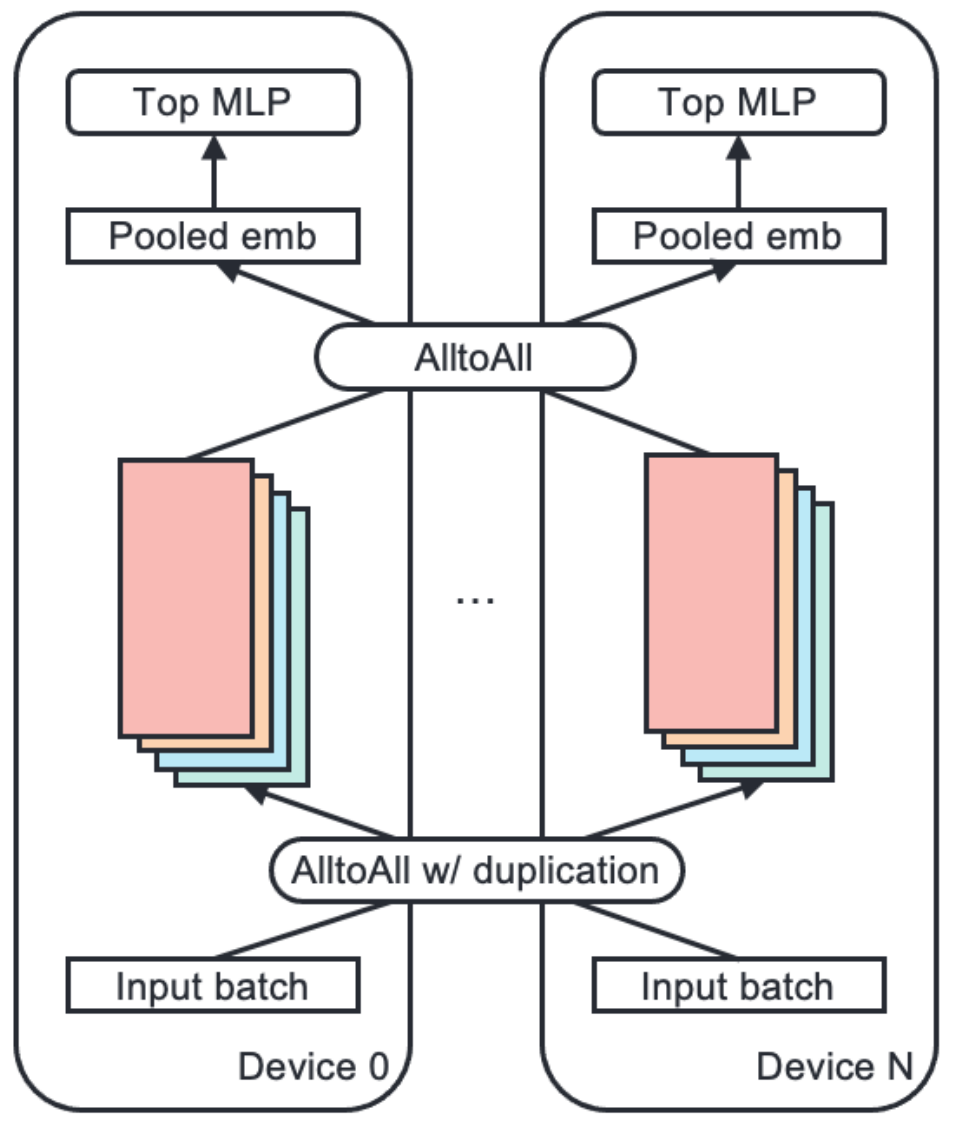}
\label{fig:sharding_cw}
}
\subfloat[Data Parallelism] {
\includegraphics[scale=0.43]{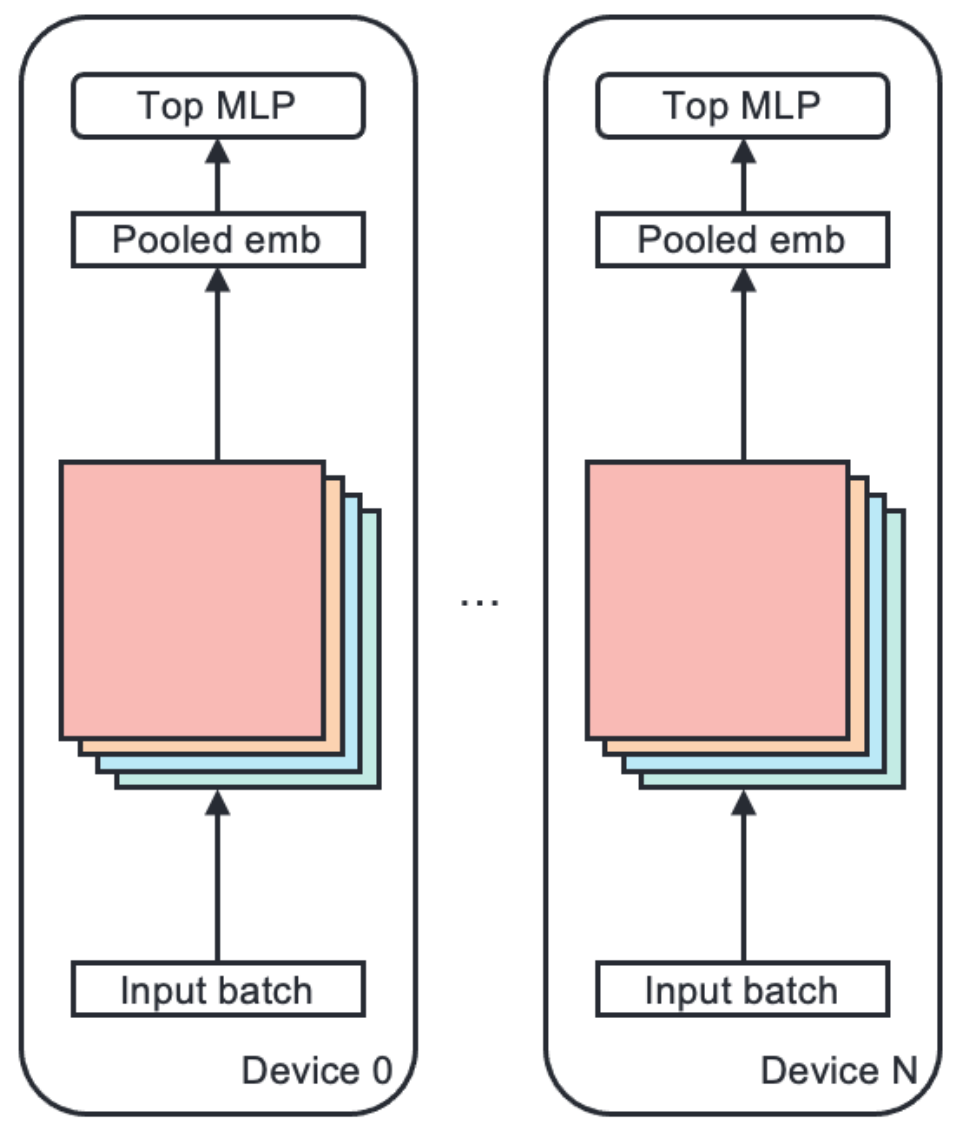}
\label{fig:sharding_dp}
}
\caption{Embedding table sharding schemes with different implications on the communication cost, load balancing and memory requirement. Bottom MLP is omitted in this figure for simplicity of illustration.}
\end{figure*}

A key component in DLRM is embedding operators, which will be defined in~\Cref{sec:embedding_ops}.
To enable high-performance training for embedding operators, it is crucial to effectively balance the workload distribution across GPUs and minimize communication costs.
We introduce {\em 4D parallelism}, which combines {\em table-wise}, {\em row-wise}, {\em column-wise}, and {\em data} parallelism for jointly optimizing the parallelization performance of embedding operators.

\paragraph{Table-wise parallelism.} 
The most straightforward parallelism scheme is partitioning and parallelizing multiple embedding tables across GPUs, as shown in~\Cref{fig:sharding_tw}.
Table-wise parallelism does not further split embedding tables, therefore this scheme requires no additional handling of embedding table input indices or pooled embedding results, leading to optimal communication efficiency. 
However, table-wise parallelism cannot handle large embedding tables that exceed the memory capacity of a single GPU, and the achieved load balance is often limited due to the skew in table sizes.

\paragraph{Row-wise parallelism.} 
This scheme parallelizes large embedding tables by rows and assigning different table shards to different trainers. Since the embedding table inputs index tables by rows, they need to be \emph{bucketized} based on the row-wise parallelism decision and distributed to the respective trainers, as illustrated in~\Cref{fig:sharding_rw}. Moreover, partial results on multiple trainers need to be reduced and then scattered to all trainers for downstream computations. This requires a \reducescatter communication pattern in the forward pass. This scheme handles large tables well and leads to better load balance. However, the communication cost scales linearly with the number of trainers.

\paragraph{Column-wise parallelism.} 
Column-wise parallelism partitions the embedding tables along the embedding dimensions (see~\Cref{fig:sharding_cw}) and treats the partitioned table with smaller embedding dimensions as individual operators. This scheme requires duplication of input indices for the partitioned tables. Compared with table-wise parallelism, it preserves the same flow and communication pattern (\alltoall). A key advantage of column-wise parallelism is enabling finer-grained parallelism, especially for large tables. However, it works well only with large embedding dimensions and increases the payload for the input indices, which have to be replicated to all nodes with the column shards. Furthermore, since the rows of column-wise sharded tables are split across different trainers, using an independent row-wise update for these tables introduces additional parameters, one for each shard of the row instead of just a single value for the entire row when using sparse optimizers (see~\Cref{subsec:fusion} for details).

\paragraph{Data parallelism.}
DLRMs tend to have a wide range of table sizes, while table-, row-, and column-wise parallelism are efficient for relatively large embedding tables prohibitive to replicate. For smaller tables, data parallelism achieves better performance, since data parallelism does not involve any communication in the forward pass (see ~\Cref{fig:sharding_dp}). 
Therefore, for small embedding tables, \sys treats embedding tables as dense parameters and replicate them across all trainers.
\alltoall is no longer needed for the pooled embeddings of data-parallel embedding tables. Instead, \allreduce is required to synchronize across all replicas. As a result, this depends on the trade-off between the cost of \alltoall of the pooled embeddings versus the cost of \allreduce on the entire table. In general, small embedding tables with fewer rows are good candidates for data parallelism. Input indices for these tables are passed through as data-parallel inputs and no longer require re-distribution.

\subsection{Parallelization Algorithms}
\sys supports applying 4D parallelism strategies at the granularity of individual embedding operators to maximize flexibility.
Practitioners can mix-and-match the above primitives to determine the best strategy to partition an embedding operator.
Additionally, \sys also supports partitioning embedding operators in a {\em recursive} manner at different levels of hardware hierarchy to further improve workload balance and hardware efficiency.
For example, the {\tt table-wise then row-wise} scheme first assigns a set of tables to a particular node, and within that node the tables are partitioned row-wise. This family of hierarchical parallelism schemes improve hardware locality by fully exploiting the fast GPU interconnects and reduce inter-node communications.

With a cost function defined for each of the above parallelism schemes, placement algorithms can be explored to minimize the cost differences between workers.
The cost function is a combination of communication overhead and load imbalance between the trainers. The communication overheads are computed using the message volume as a representative metric, with higher message volumes corresponding to higher costs. This is largely accurate in capturing the throughput costs and for latency measured values are incorporated as a fixed additive cost. We estimate the load imbalance by using the embedding access size per trainer, which can be approximated as the number of embedding tables per trainer $\times$ the global batch size $\times$ average number of indices per sample $\times$ embedding dimension . The combination of both costs gives us a reasonable estimate for communication and load imbalance. Further we introduce scalar weight for each of the individual costs, which can be tuned based on different system specs to get more accurate estimations.

We implement and evaluate two polynomial time heuristics as a proof of concept. The first one is a simple greedy heuristic that sorts the costs of available schemes in a descending order and allocates the largest shard first, one per worker. Then, the greedy algorithm iterates through all remaining shards and assigns the top cost to the node with the smallest sum of costs. A second heuristic is the largest differencing method (also known as the Karmarker–Karp algorithm~\cite{ldm}). The main idea is to take the two largest numbers from the input and replace them by their difference. It directly reduces the difference of sums and generally outperforms the greedy heuristic.

\subsection{Pipelining}
\label{sec:pipeline}

Although using GPUs as the main compute resource offers limited pipelining opportunities within model evaluation, we improve GPU utilization by pipelining inter-batch data movement and overlapping communication with computation.

When batch $i$ is being evaluated, the same GPUs can start receiving and distributing batch $i+1$ using a separate stream. To minimize the interference, we overlap the input \alltoall of batch $i+1$ with the forward propagation of top MLP of batch $i$ where no communication is involved. In addition, we overlap the pooled embedding \alltoall with the forward propagation of bottom MLP to hide latency.

\section{Embedding Optimizations}
\label{sec:embedding_ops}
Optimizing the runtime performance of DLRM's embedding operators (see~\Cref{sec:background}) requires addressing two key challenges.
First, the forward processing, backward propagation, and gradient updates for the embedding operators require launching thousands of GPU kernels in each training iteration, introducing significant GPU kernel launch overhead.
Second, some embedding operators may include up to billions of parameters and do not fit on the device memory of a single GPU. 

We introduce three novel techniques to reduce the computational cost and memory requirement of embedding operators.
First, we introduce a {\em hybrid kernel fusion} technique to minimize the CUDA kernel launch overhead and allow each GPU worker to only launch two kernels (i.e., one for forward and one for back propagation and parameter update).
Second, for parallelizing the computation of the embedding operators, we propose {\em column-wise parallelism} and {\em row-wise parallelism} in addition to data and model parallelism.
The combinations of these four parallelism dimensions enable \sys to support embedding tables with up to trillions of parameters.
Finally, \sys exploits a series of memory saving techniques that leverage the memory hierarchy of the \zionex platform to ensure sufficient memory capacity for DLRM.

\begin{figure}
\centering
\subfloat[Fusing multiple embedding tables] {
\includegraphics[width=\linewidth]{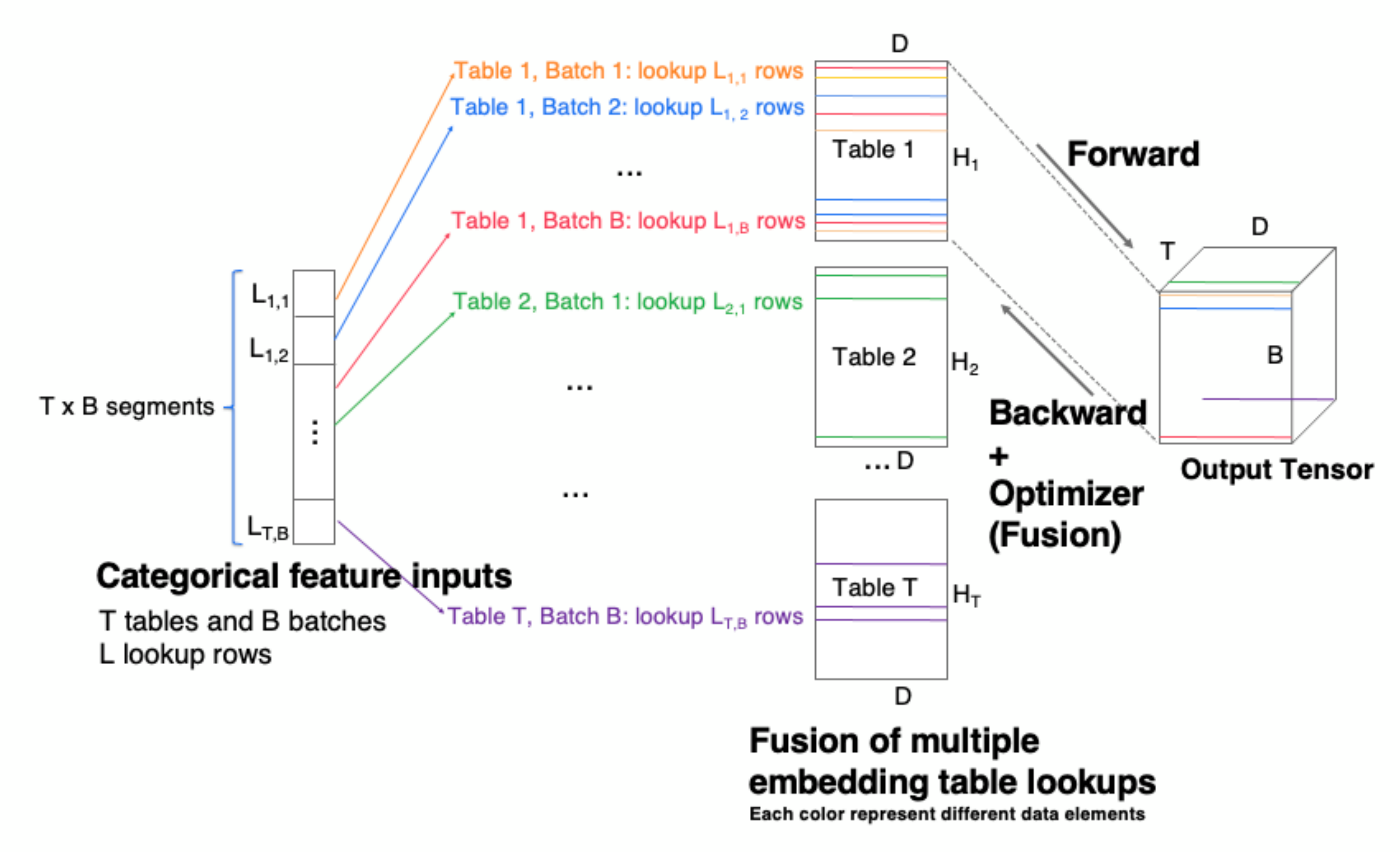}
\label{fig:emb}
}
\\
\subfloat[Fusing embedding backward and sparse optimizer] {
\includegraphics[scale=0.36]{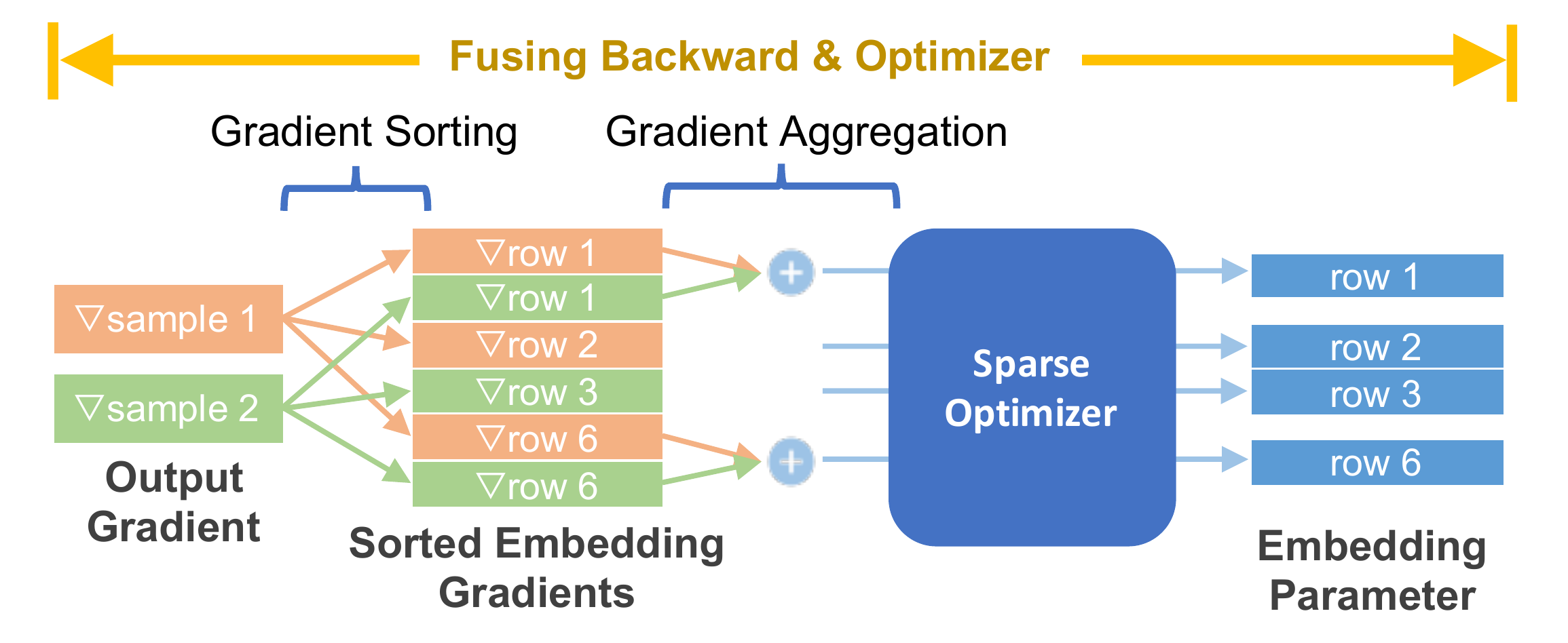}
\label{fig:embedding-fused}
}
\caption{Embedding operator optimizations}
\end{figure}

\subsection{Kernel Fusion}
\label{subsec:fusion}
\sys uses a {\em hybrid kernel fusion} mechanism to minimize the CUDA kernel launch overhead for performing embedding computations in a training iteration.
First, instead of applying a separate embedding lookup for each embedding table, \sys fuses multiple embedding lookups on the same GPU into a single CUDA kernel (Figure~\ref{fig:emb}), which improves the parallelism and bandwidth utilization and reduces the overhead of launching multiple CUDA kernels on GPUs.

Second, \sys also fuses the backward pass with the sparse optimizer to further reduce kernel launch overhead and avoid materializing gradients to the embedding tables.
The key challenge of such fusion is avoiding potential race-condition across gradient updates from different training samples and handling non-linearity in advanced optimizers such as AdaGrad~\cite{adagrad}, LAMB~\cite{lamb}, and Adam~\cite{adam}.
For example, both sample 1 and 2 in~\Cref{fig:embedding} contribute to the gradients of the embedding vector 1 and 6. Directly sending these gradients to a non-linear sparse optimizer without aggregation would result in incorrect updates to the embedding tables.

To guarantee correctness while maximizing performance, \sys applies {\em gradient sorting} by rows so that gradients to the same embedding rows are processed by a single CUDA thread block, as shown in~\Cref{fig:embedding-fused}. {\em Gradient aggregation} is subsequently applied within each CUDA thread block using much faster but smaller GPU shared memory.

\sys's hybrid fusion technique for embedding operators lead to three performance benefits.
First, \sys reduces the memory requirement for embedding operators by avoiding allocating GPU device memory for embedding gradients.
Second, the memory accesses to GPU device memory are minimized by using GPU shared memory to save intermediate embedding gradients.
Finally, kernel fusion improves the overall performance of embedding computations by up to 7$\times$ compared to a native implementation. The optimized embedding operator implementations are open sourced as part of the {\tt FBGEMM} library\footnote{PyToch FBGEMM\_GPU library: \url{https://github.com/pytorch/FBGEMM/tree/master/fbgemm_gpu}} and integrated with PyTorch.

\subsection{Managing Memory Hierarchy}

For DLRMs with up to trillions of parameters, the embedding tables are too large to entirely fit on a single GPU. We leverage multiple levels of memory hierarchy of the \zionex platform, including HBM, DRAM and SSDs in additional to scaling out to multiple nodes for increasing aggregate capacity, to ensure sufficient memory for the models, with the faster memory serving as a software cache of the subsequent layer.
\sys's hierarchical memory management strategy is specifically useful for online training of DLRMs, which warrants using fewer nodes for training original large models, due to lower throughput requirements, as outlined in Sec.~\ref{sec:background}.
One approach to managing memory hierarchy is CUDA's unified memory (UVM)~\cite{unified_memory}, which provides a single memory address space for different memory types and automatically replaces and evicts unused pages.
However, random table lookups in embedding operators requires caching and replacing unused parameters at the granularity of individual embedding rows, which makes using UVM as-is insufficient for DLRM. Necessitating additional handling of the look-up to ensure performance is not bound by the frequent host to device transfers.
Instead, \sys uses a customized 32-way set-associative software cache~\cite{high_prec_cache} using least recently used (LRU) or least frequently used (LFU) cache replacement policies, where the associativity matches the warp size of GPUs.
This enables fine grain control of caching and replacement, allowing it to be tuned for target model characteristics.
Note that UVM is bounded by PCIe bandwidth, while \sys's software cache can bridge the gap for the bandwidth between PCIe and HBM (~50$\times$ difference).
The software cache improves the end-to-end performance of DLRM workloads by approximately 15\% compared to UVM.

To further reduce the memory requirement of embedding operators, \sys also employs a variety of compression techniques introduced in prior work, such as a row-wise sparse optimizer~\cite{gupta2014training,sc20_kraken}, low/mixed-precision training using a high-precision cache backed by low precision embedding tables~\cite{lowprecision}, and advanced factorization techniques~\cite{koren2009matrix}.


The row-wise sparse AdaGrad was first introduced in~\cite{gupta2014training}, and then further elaborated in~\cite{sc20_kraken}.
In the row-wise sparse AdaGrad, each element of the moment estimation is applied to the entire embedding row. For each row it is a single scaling factor that is updated by adding the average squared sum of gradients across the row.
In this way, we keep the momentum as a 1D tensor with $ H $  elements instead of $ H \times D $ 2D tensor, where $H$ and $D$ are the number of rows and the number of elements per row in an embedding table, respectively.


\section{\zionex: Hardware Platform Design}

\begin{figure}
    \centering
    \subfloat[\zion]{
        \includegraphics[width=\linewidth]{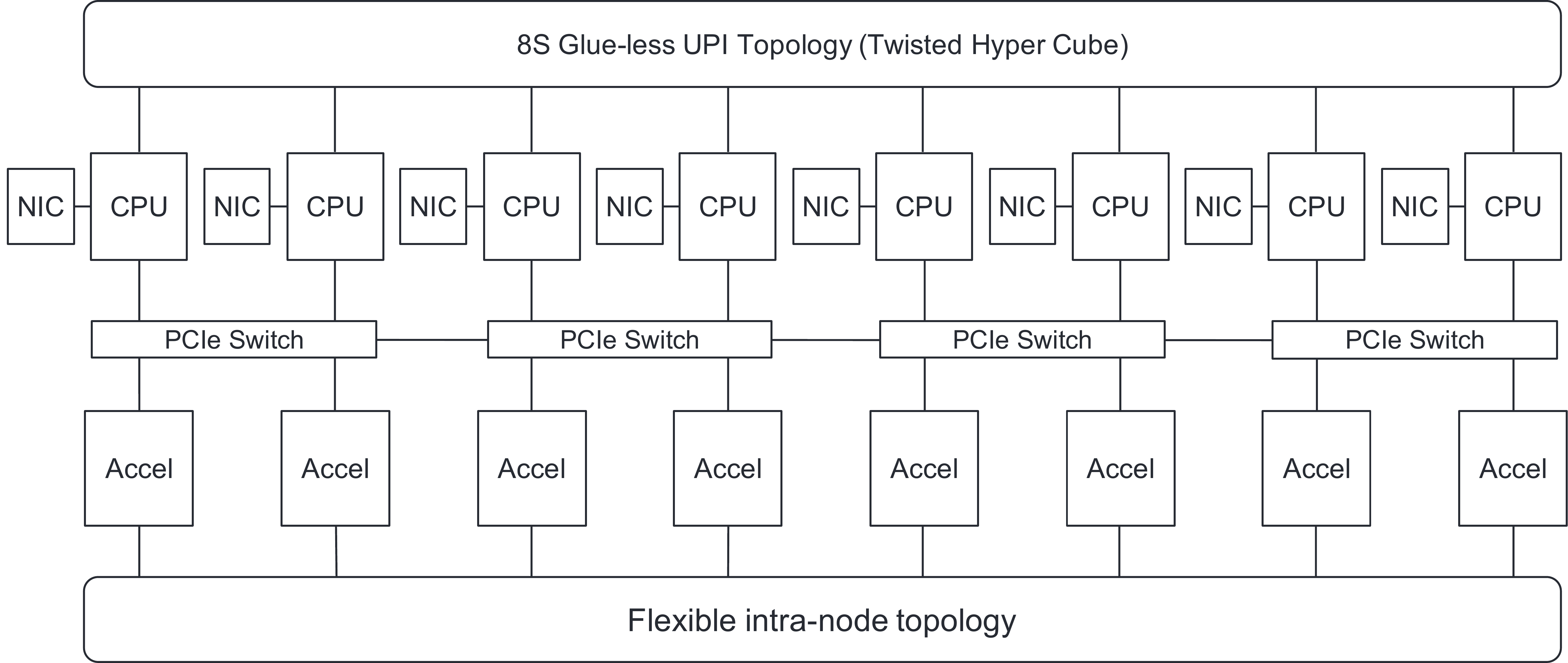}
        \label{fig:zion}
    }
    \\
    \subfloat[\zionex]{
        \includegraphics[width=\linewidth]{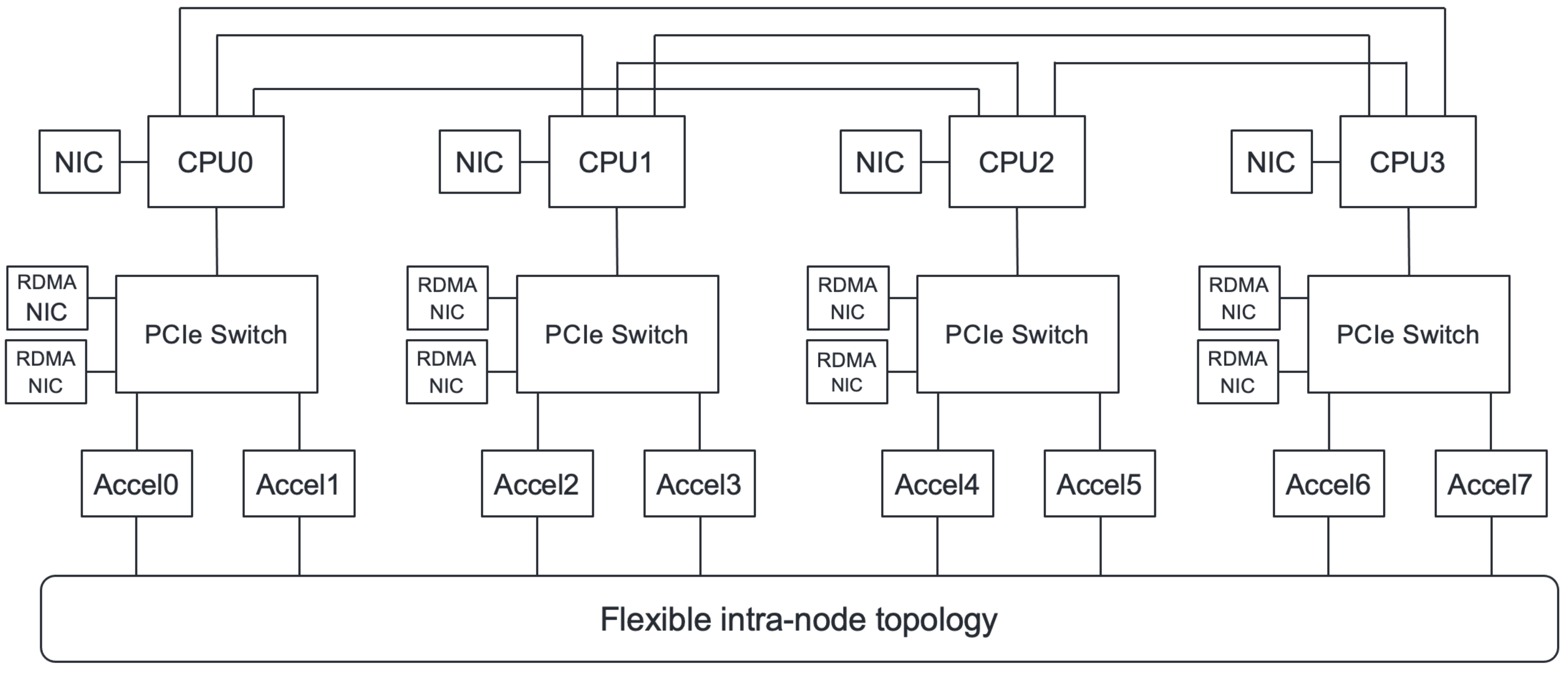}
        \label{fig:zionex}
    }
    \\
    \subfloat[Overall training architecture.]{
    \label{fig:system}
    \includegraphics[width=\linewidth]{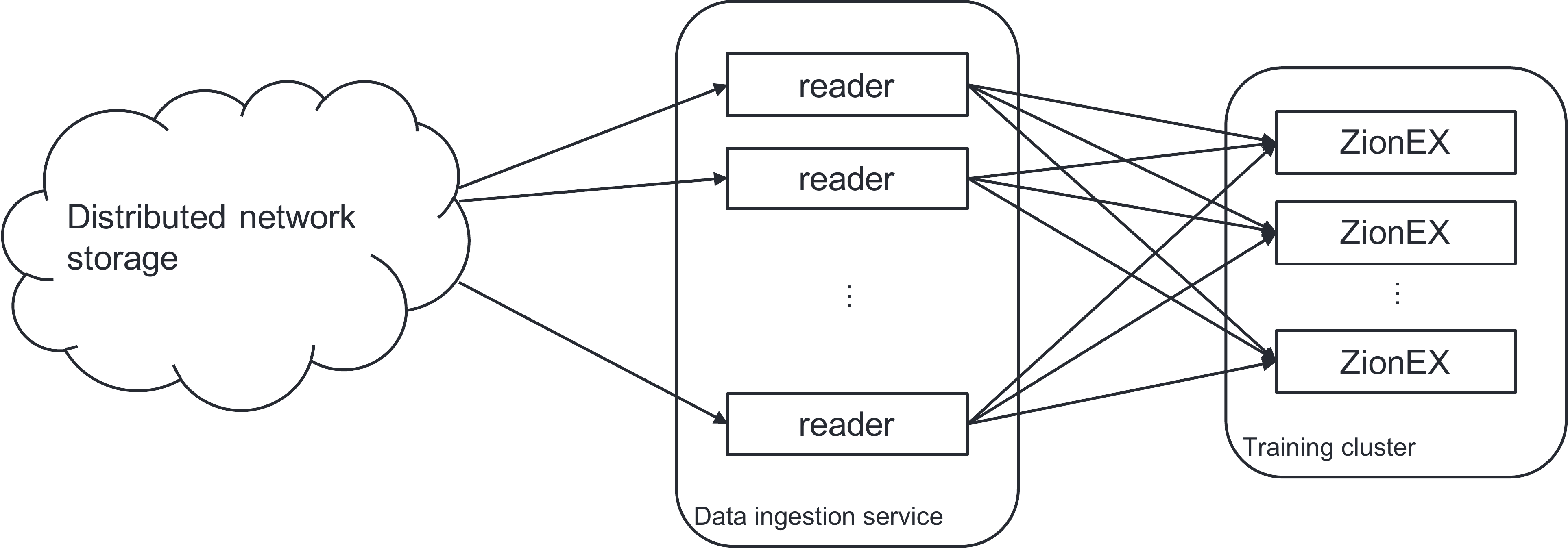}
    }
    \caption{The system architectures of \zion, \zionex platforms and the overall training system. 
    }
\end{figure}

We begin by describing the limitations of our previous hardware platform for DLRM in~\Cref{subsec:zion}. ~\Cref{subsec:zionex} introduces \zionex, a new hardware platform for DLRM. We also outline the design principles used in the development of \zionex.

\subsection{Previous Platform: \zion}
\label{subsec:zion}
\zion~\cite{zion_ocp} introduced in 2019 was our previous work aimed as a high-performance hardware platform for training DLRMs.
While \zion offers significantly improved capabilities at single-node level, it falls short as a distributed platform not being extensible to meet the rapidly growing DLRM training requirements.
We critically appraise its limitations, but other platforms based on a similar design share the same limitations; we discuss those platforms in~\Cref{sec:related}. 

\Cref{fig:zion} shows the architecture of a \zion node, which has 8 CPU sockets with 1.5 TB memory, 8 GPUs, and 8 network interface cards (NICs).
It provides a powerful heterogeneous {\em super node} design for training DLRM by (1) offloading compute heavy layers of DLRM (e.g., MLPs) onto GPUs and (2) leveraging CPUs for large embedding operators on the relatively cheaper DRAM instead of HBM for accommodating TB-scale DLRMs on a single node.

However, this heterogeneous design introduces a number of challenges to software design and performance.
For example, it's critical to balance the workload on CPUs and GPUs to ensure maximum overlap. This requires elaborate pipelining between CPUs and GPUs and partitioning DLRM into fine-grained tasks using an accurate cost model.
In addition, heterogeneous training of DLRM also introduces non-trivial runtime overheads, such as increased data transfers between CPUs and GPUs and inter-socket communication.

Finally, a critical missing component in \zion is that each NIC is directly attached to a CPU.
As a result, all of the inter-node communications (e.g., gradient synchronization and tensor transformation) necessitate CPU intervention and additional GPU-CPU transfers.
Furthermore these NICs are connected to the common shared data-center network infrastructure, which introduces overheads and interference from network congestion, and are constrained to use more data center-friendly topologies, protocols (TCP/IP) which are sub-optimal for distributed training. 
Although each \zion node is equipped with 8x 100Gbps NIC bandwidth, in reality we found it is very difficult to scale out to multiple nodes due to networking overheads.
With today's increasing demand on modeling size of DLRMs, \zion is not able to scale well and fully utilize the powerful hardware resources.


\subsection{\zionex}
\label{subsec:zionex}

To address these shortcomings, we introduce \zionex, which we have designed to be more scalable than the previous \zion platform with improved network capabilities, while retaining its flexibility and core advantages, such as the OAM form factor, modular design~\cite{zion_ocp, zion}, and flexible intra-node accelerator fabric~\cite{zion_nocs}. With all of these improvements \zionex bring about orders of magnitude higher capability both in terms of supporting increased model complexity and higher training performance. This is best illustrated by comparing the product of maximal model complexity (in terms of FLOPS/sample) supported by each platform and achieved training throughput, which can be seen as normalized effective performance. For \zionex with achieving a throughput of 1.2 MQPS for a model with 638 MFLOPS/sample (from Table.\ref{tab:models}), this translates into a effective performance of 766 TFLOPS/s, with additional headroom to go up to several PETAFLOPS/s. Whereas for \zion, the maximal model complexity that could be supported was less half of that on \zionex ($\approx 250 MFLOPS/sample$) and with much lower throughput ($\approx 0.25 MQPS$)\cite{acun2021hpca, naumov2020deep}, thereby leading to more than $10\times$ lower max achievable effective performance of only 63 TFLOPS/s.
\Cref{fig:zionex} shows the overall system architecture. We briefly highlight \zionex's core design principles:

\paragraph{Scalability.}
Both \zion and \zionex support heterogeneous training of DLRM, but the most striking difference is that \zionex is designed with sufficient scale-up and scale-out network capabilities.
As shown in~\Cref{fig:zionex}, \zionex employs a dedicated \emph{RDMA over Converged Ethernet} (RoCE) NIC for each of the GPUs connected via PCIe switches to allow for a dedicated inter-node connectivity (isolated from common data-center network) and importantly support more efficient RDMA/GPUDirect communication protocols \cite{naumov2020deep}.
These \zionex nodes can be connected with a dedicated backend network to form a cluster for distributed scalable training. The extensible design of \zionex allows for scaling the backend network to interconnect many thousands of nodes, forming a \emph{data-center scale AI training cluster}.

\paragraph{High Performance.}
As a scale-out solution, we offload the entire DLRM to GPUs to fully leverage the massive parallelism and high memory bandwidth to accelerate MLPs and embedding computations.
To transfer tensors and synchronize gradients, each GPU can communicate directly with GPUs on a different node through the dedicated low-latency high-bandwidth RoCE NIC, without involving host CPUs.
In addition, \zionex also has a frontend NIC connected to each CPU. Data ingestion goes through the regular frontend network and PCIe, without interfering with activations or gradients. The host CPUs are only used to setup input batches and marshal the training process.

\paragraph{Capability.}
With \zionex we ensure that the platform is compatible with existing infrastructure and can be widely deployed within our data-centers, without causing major disruptions. This is critical for being able to effectively leverage the capability of the platform and make it readily available to across variety of applications and uses-cases. 
We achieve this by making the \zionex platform compliant with the standard Open Rack specifications~\cite{orv2}
, which covers the compatibility with other infrastructure components such as power, cooling, mechanicals and cabling. Furthermore designing the platform to be modular and relying on open standards based technologies, for instance - the ethernet based network fabric for high-performance scale out solution.

Fig.~\ref{fig:system} shows the overall training platform, along with the dis-aggregated data-ingestion service. This supports streaming input data from a network store such as Tectonic \cite{fbwarmstorage} and perform light-weight data pre-processing operations in a distributed fashion. So that the data-ingestion is not a bottleneck for the end-to-end training and to ensure sufficient throughput in feeding \zionex trainers.
\section{Implementation}
We detail the implementation of high-performance scalable training for DLRMs described above. 
We built a high-performance training software stack for DLRMs using PyTorch~\cite{pytorch}, with efficient CUDA implementation for most deep learning operators via the {\tt ATen} library, and automatic handling of parameter replication and gradient synchronization with overlapped back-propagation and \allreduce via the PyTorch {\tt DistributedDataParallel} library~\cite{pytorchddp}. 
We have enabled the following components for efficient DLRM training.
\subsection{Data ingestion} 
\label{sec:data}

Data ingestion is a key component to ensure end-to-end training performance especially for DLRMs, which typically process through order(s) of magnitude larger amount of data than other typical DNN models. We observe that data ingestion, if left unoptimized, can incur significant latency and introduce non-trivial overheads to pipelining.

Originally designed for a distributed asynchronous CPU setup, our readers and data pre-processing module stores the offsets and indices\footnote{Please refer to the interface of {\tt nn.EmbeddingBag} \url{https://pytorch.org/docs/stable/generated/torch.nn.EmbeddingBag.html}} of each sparse feature in separate tensors per embedding table. As a result, a DLRM with hundreds of embedding tables can easily get a thousand input tensors per iteration, which translates into significant overheads from \emph{CPU $\leftrightarrow$ GPU} transfers and was one of the key bottlenecks for the previous Zion platform as detailed in Sec.~\ref{sec:background}. 

To overcome this practical challenge, we co-designed the data pre-processing module to use a combined format where lengths rather than offsets are used and inputs to different embedding tables are simply concatenated. The benefits of using the combined format are two-fold: (1) it optimizes CPU-GPU transfer by consolidating small transfers; (2) it can be directly consumed by the embedding kernel without additional layout transformations. We further optimized input data transfer by using pinned memory to avoid the extra copy.

With the combined format, we developed a module to efficiently distribute embedding table inputs based on the sharding strategy. In the case of table-wise sharding (shown in Fig.~\ref{fig:sharding_tw}), an \alltoall is needed to distribute the global batch for local tables to each worker. Since the size of indices is dependent on the values of the lengths, the communication is actually implemented as an \alltoall for lengths followed by an \alltoall for indices. In a setup with $W$ workers, $T$ local tables and $B$ local batch size, this gives us indices in the order of $(W, T, B)$, which needs to be further permuted to $(T, W, B)$ for embedding kernel consumption. We have developed custom GPU kernels for permute, bucketize and replicate to achieve maximum throughput on embedding input indices distribution for table-wise, row-wise and column-wise sharding schemes. Checkpointing the model has similar challenges, requiring to be sufficiently frequency  be able to write-out such larger model whilst not becoming an overhead for training, as outlined in this recent paper~\cite{checknrun}.

\subsection{Communication Primitives}
High-performance collective communication is key to performant and scalable DLRM training.
PyTorch provides the Process Group (PG) interface for collectives - an abstract platform / collectives library agnostic API. DLRM uses this API directly (for Alltoall) or indirectly via DDP (for Allreduce) \cite{pytorchddp}. We use the \emph{NVIDIA's Collective Communication Library} (NCCL) as our primary collective communication library since it efficiently uses RDMA and NVLINK for best performance. We extended PyTorch NCCL process group implementation to support Alltoall/Alltoallv collectives using NCCL Send/Recv primitives (requires NCCL 2.7.3 or later). 
\section{Evaluation}
We provide results for end-to-end training of production models, operator-wise performance breakdown.

\subsection{Experimental Setup}


\begin{table}[t]
\small
\centering
\caption{\zionex per-node system specification.}
\label{tab:zionex}
\begin{tabular}{ll}
\toprule
Compute (TFLOPS)             & 156 (FP32) / 1248 (TF32) / 2496 (FP16,BF16)      \\ \midrule
HBM                          & 320 GB, 12.4 TB/s            \\ \midrule
DDR                          & 1.5 TB, 320 GB/s             \\ \midrule
Scale-up bandwidth           & 2.4 TB/s (uni-directional)   \\ \midrule
Scale-out bandwidth          & 1600 Gbps (uni-directional)  \\ \midrule
Host NW                      & 4 $\times$ 100 Gbps          \\
\bottomrule
\end{tabular}
\end{table}



Table~\ref{tab:zionex} summarizes the aggregated capabilities of a single \zionex node with 8 NVIDIA A100 GPUs. The 8 GPUs in a node provide a total 320 GB HBM with 12.4 TB/s aggregated memory bandwidth. The 4-socket CPUs provide 1.5 TB memory with 320 GB/s bandwidth. On network capabilities, the GPUs are interconnected with high-bandwidth NVLink for intra-node GPU communication, and each GPU has a dedicated 200 Gbps RoCE NIC for inter-node communication. We use a cluster of 16 \zionex nodes in the experiments with 5TB total HBM capacity.



\begin{table}[b]
\small
\caption{Target models configuration}
\label{tab:models}
\begin{tabular}{l|ccc}
\toprule
Model & model-A & model-F & model-I  \\ \toprule
Num parameters & 793B & 12T & 332B \\ \midrule
MFLOPS per sample & 638 & 5 & 60 \\ \midrule
Num of emb tables & $\approx~1000s$ & $\approx~10s$ & $\approx~100s$ \\ \midrule
Embedding table dims & [4, 384] & [256, 256] & [92, 92]\\
(range [min, max], avg) & avg: 93 & avg: 256 & avg: 92 \\ \midrule
Avg pooling size & 15 & 20 & 70 \\ \midrule
Num MLP layers & 20 & 7 & 43 \\ \midrule
Avg MLP size & 3375 & 490 & 682 \\ \bottomrule
Target local batch size & 512 & 512 & 2048 \\ \midrule
Achieved QPS & 1.2M & 1.7M & 3.4M \\ \bottomrule

\end{tabular}%
\end{table}

\subsection{End-to-End Training}



We report results on three DLRMs deployed in production for different tasks, including click through rate (CTR) prediction, ranking, and engagement.
\Cref{tab:models} lists high-level characteristics of these candidate models. Model-A represents large and complex DLRMs that stress \sys's compute capability and communication bandwidth, using significantly higher FLOPS per sample and a large number of embeddings.
Model-F presents a different practical challenge where despite having low FLOPS per sample and a small number of embedding tables, it has a single massive table that cannot fit in the device memory of a single GPU.
Finally, Model-I represents moderate scale DLRMs stressing memory bandwidth with high average embedding pooling sizes.
These target models are trained on up to 16 \zionex nodes (128 GPUs) in the cluster. The model qualities are evaluated in normalized entropy~\cite{ne}, and the training throughput is measured in queries per second (QPS).

\begin{figure}[t]
    \centering
    \includegraphics[width=\linewidth]{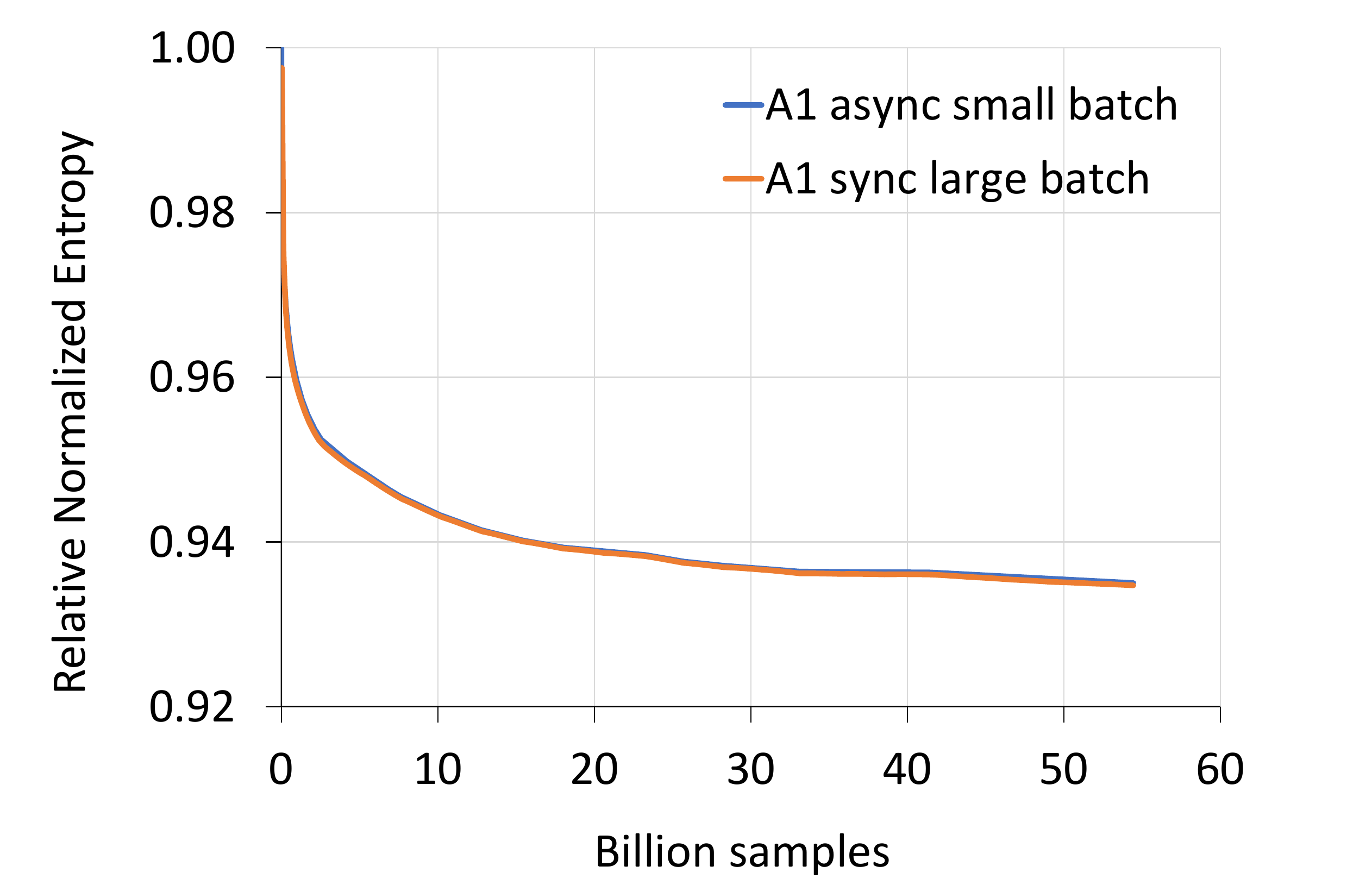}
    \caption{Training quality comparison between asynchronous small batch on a distributed CPU platform and synchronous large batch on the proposed platform, measured in relative normalized entropy~\cite{ne}.}
    \label{fig:ne}
\end{figure}

First, we use model-A to demonstrate the training quality, since it can also be trained on a distributed CPU platform. As shown in~\Cref{fig:ne}, despite using significantly larger batch size (64K vs. \textasciitilde 150), 
synchronous large batch training on \zionex provides on-par or better model quality (both using tuned hyperparameters).
With the same configuration, \sys achieves 1.2 MQPS 
using 128 GPUs on 16 nodes, a 40$\times$ speedup compared to our previous generation distributed CPU asynchronous training platform using 45 parameter servers and 15 trainers. While previous solution was unable to scale out further without hurting training quality, fully synchronous training on \zionex allows scaling beyond 16 nodes with even larger batch sizes. 


\subsection{Scaling Performance}
\label{sec:scaling}
\begin{figure}[t]
    \centering
    \includegraphics[width=0.9\linewidth]{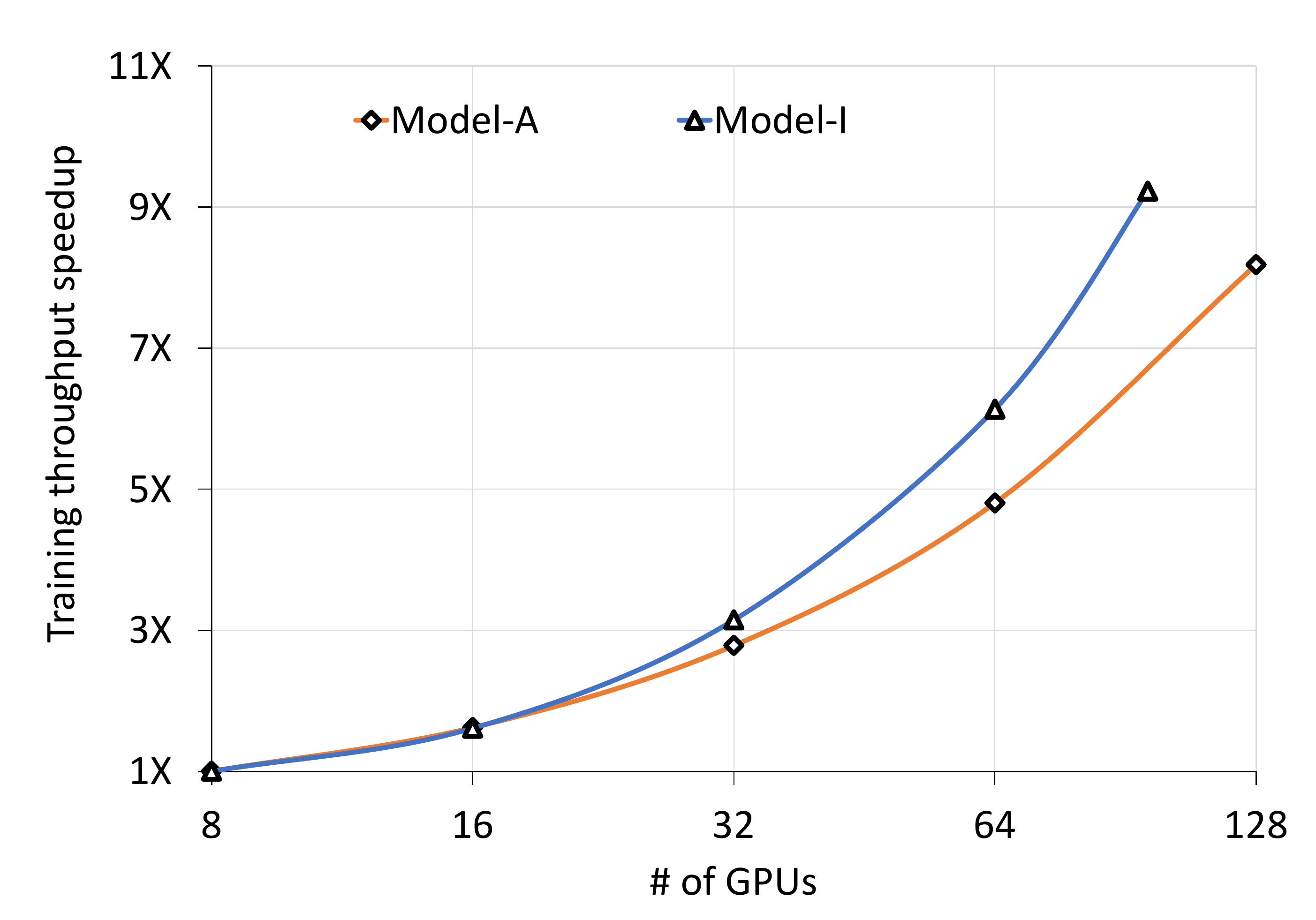}
    \caption{Training throughput scaling for model-A and model-I, relative to 8 GPUs (1 node).}
    \label{fig:a_scaling}
\end{figure}

\Cref{fig:a_scaling} shows the normalized training throughput of model-A and model-I using up to 16 nodes, while keeping the per-GPU batch size constant. 
While the workload of data-parallel training remains the same with the scaling, the numbers of embedding tables per GPU reduces with scaling due to model-parallelism. 
For the same reason, however, each GPU processes the entire global minibatch for each of its local tables and this increases commensurately with scale and compensating for the reduced tables, making this still a weak scaling experiment. To run on smaller node counts, we reduce the embedding table cardinality and hash inputs to be within the reduced number of rows. This \emph{shrunk version of the model} effectively reduces the model sizes with minimal/no impact on the performance characteristics, hence is used for studying scaling performance.

As seen from the figure, on larger node counts, the scaling efficiency is around 50\% for model-A and around 75\% for model-I. While model-A and model-I come very close in terms of effective FLOPS and memory requirements after considering the target local batch size, model-A has larger fully exposed \alltoall latency. This is because more embedding tables increase \alltoall payload, and mixed dimensions make it more difficult to balance embedding computations and \alltoall communications at the same time. As a consequence, model-A suffers more from reduced \alltoall efficiency when scaling out.

\begin{figure}
    \centering
    \includegraphics[width=\linewidth]{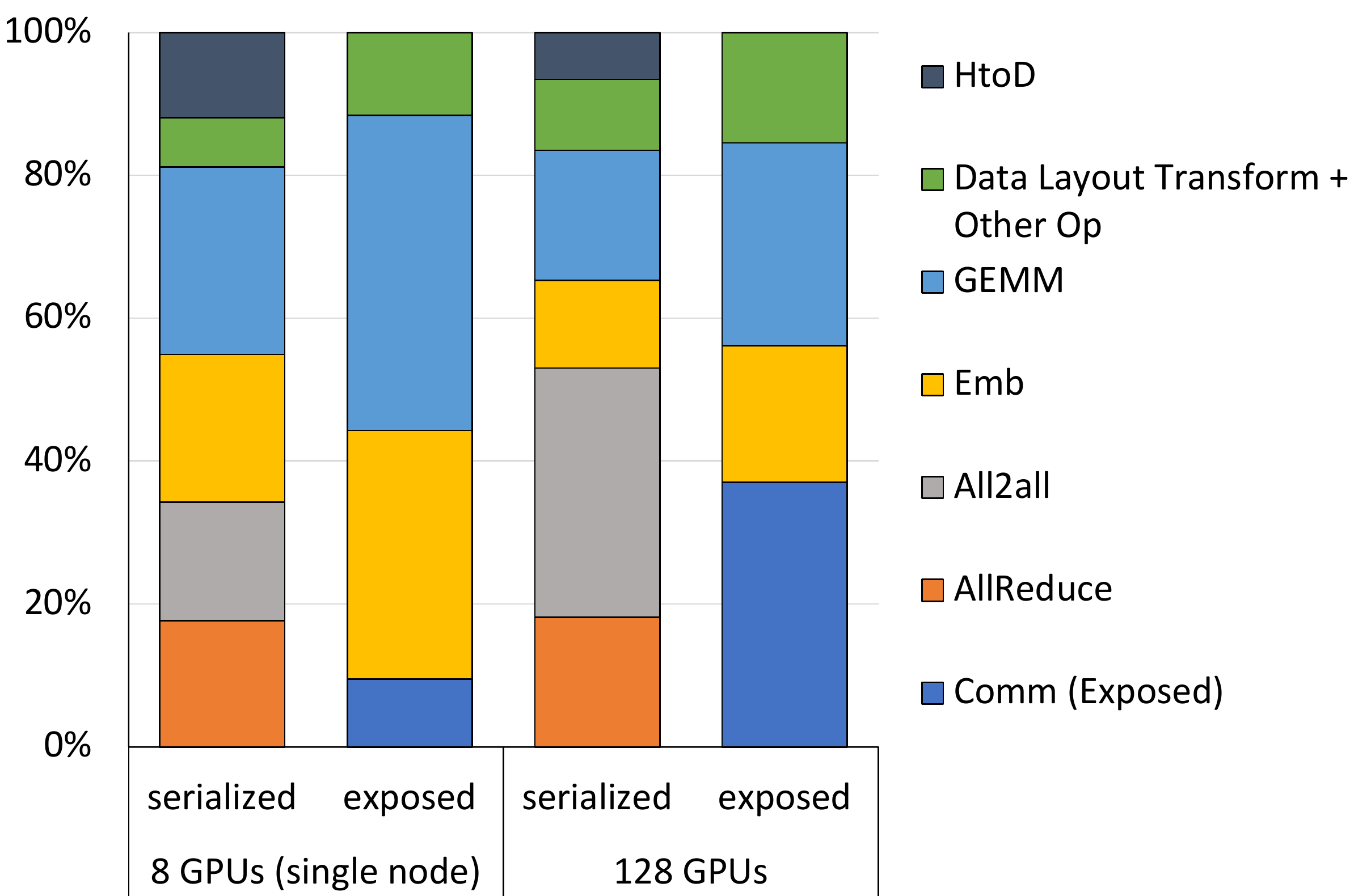}
    \caption{Model-A (with local batch size per GPU = 512) results and dominant operator time breakdown (serialized and exposed time) per GPU, after optimizations.}
    \label{fig:a2_breakdown}
\end{figure}

To better understand the scaling performance, we provide a breakdown of serialized and exposed training iteration latency of model-A in Figure~\ref{fig:a2_breakdown}. Comparing between serialized and exposed latency, the CPU to GPU transfer (i.e., HtoD) is completely hidden, and the exposed communication latency is much less than serialized \alltoall and \allreduce latency combined.
This demonstrates the effectiveness of \sys's pipelining optimization to overlap communications with computations (see~\Cref{sec:pipeline}).

As node count increases, we observe increased \alltoall and \allreduce latencies. Since most \alltoall communications are on the critical path, increased \alltoall cost has a direct impact on the exposed communication and overall training latency. 
While \allreduce is mostly hidden on up to 16 nodes, the increased \allreduce latency and unchanged computation latency signifies that \allreduce can become the bottleneck once the slack in backward pass is completely used up with higher node counts and/or faster computation.

\subsection{Training Throughput Optimizations}
\begin{figure}[t]
    \centering
    \includegraphics[width=0.95\linewidth]{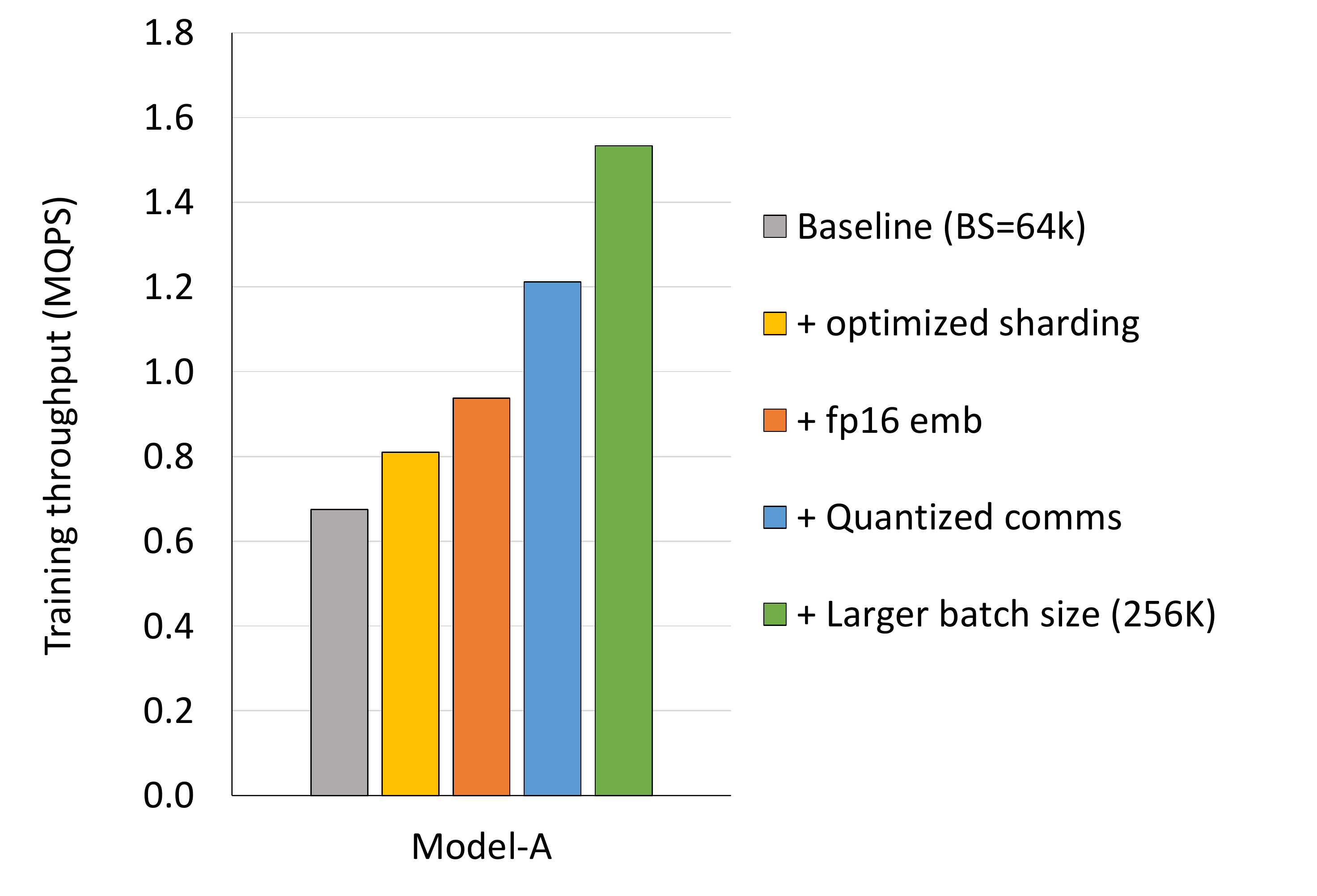}
    \caption{Training throughput improvements enabled by optimized sharding, reduced precision embeddings, quantized communications and larger batch sizes.}
    \label{fig:a2_perf_impr}
\end{figure}

Using model-A as a case study, we detail the various optimizations and their contributions in achieving up to 1.5 MQPS, shown in~\Cref{fig:a2_perf_impr}. Further, we use the performance roofline modeling methodology described in ~\href{app:B}{Appendix-B} to establish the upper bound of achievable performance and confirm that reported throughout is within $~15\%$ theoretical estimates. The baseline performance for model-A on 128 GPUs is below 700 KQPS. Further profiling reveals large disparities on embedding lookup latency between different GPUs, signifying severe load imbalance. This is mitigated using a combination of table-wise, column-wise, and data parallelism for the $\approx 1000s$ of embedding tables to partition them across 128 GPUs. Note that even though column-wise parallelism introduces additional cost to its input \alltoall, the benefit from better load-balance outweighs the overheads and results in overall QPS improvement by 20\%. However, the scaling efficiency is still about 30\% lower than ideal linear scaling.

As discussed previously, the two major issues limiting scaling efficiency are: (1) load imbalance and (2) increased \alltoall latency. For model-A, further balancing the load using only HBM is particularly challenging because the model size in TF32 comes close to the 5TB aggregated HBM capacity on 128 GPUs. After discounting for memory reserved by PyTorch framework and NCCL on each rank, \sys has little room to explore placement strategies. 
To mitigate this issue, we use lower precision (FP16) embedding tables~\cite{zhang2018training}, reducing the model size by a factor of 2. While this alone does not provide direct throughput benefit, \sys can leverage the head room to strike a better balance. As a consequence, the training throughput is increased by another 20\% due to improved load balancing.

Next, to address the increased \alltoall latency, we incorporate quantized collective communications proposed in~\cite{qcomm}, which directly reduce the communication volume. For model-A, we validate that using {\tt FP16} in forward \alltoall and {\tt BF16} in backward \alltoall provides almost 30\% speedup without any training quality loss.

Lastly, we increase the global batch size from 64K to 256K. This directly increases activation sizes, which helps saturate GPUs and communication bandwidth better, while being complimentary to all other optimizations.
With appropriately tuned optimizer/hyper-parameters, we are able to achieve on-par training quality, however more comprehensive experimentation is warranted since large batch training of DLRMs is not as well studied and will be part of future work.
Collectively, these techniques unlock an 87\% improvement on training throughput compared to {\tt TF32} training with a 64K global batch size.

\subsection{Model Capacity Limitation Study}
\label{sec:f1}

We use model-F as an example to push the model capacity on the prototype system. Unlike model-A or model-I, efficiently training model-F presents 2 different challenges. First, with 12T parameters, model-F can easily require up to 96TB of memory using a naive training approach, far exceeding the total memory available on a 16-node cluster\footnote{Considering FP32 precision and doubled size for optimizer states $12e12 \times 4 \times 2 = 96e12$. The prototype cluster has in total 4TB HBM and 24TB DRAM}. Second, the model has only a few massive embedding tables with \textasciitilde 10B rows and 256 columns, each requiring multi-node worth of GPU and host memory to train.

To fit the model onto 16 nodes, we first apply row-wise sparse AdaGrad optimizer to embedding tables which reduces optimizer states from per element to per embedding row. Then we use FP16 precision on embedding tables~\cite{zhang2018training}. These two optimizations collectively bring model memory footprint from 96TB down to 24TB, just fitting under the 4TB HBM + 24TB DRAM memory hierarchy. On the massive embedding tables, we enable row-wise sharding to distribute the tables to multiple nodes and adjust the training flow to use \alltoall with bucketization and \reducescatter as shown in Figure~\ref{fig:sharding_rw}. With UVM enabled and HBM used as a cache, we are able to train model-F with throughput as high as 1.7 MQPS, demonstrating capability of our HW/SW co-designed solution to push beyond the current state-of-the-art.
\section {Related work}
\label{sec:related}
 
Researchers have proposed various system-level innovations to tackle the challenges from extremely large models. DeepSpeed~\cite{deepspeed} fully shards model parameters, gradients and optimizer states across all nodes, and reconstructs necessary states on the fly using checkpoint partitioning and rematerialization~\cite{dtr, checkmate} to drastically reduce memory usage. GShard~\cite{lepikhin2020gshard} trains a massive translation model with mixture of experts, sharded across accelerators through annotation of parallelization strategy at tensor level.
FlexFlow~\cite{flexflow} uses automatic search to discover the best operator parallelization strategy in the graph. Building on this direction of auto-parallelization, these recent papers \cite{tarnawski2020efficient, deviceplacementgoogle1} use optimal synthesis and reinforcement learning to find optimized device placement to further improve parallelism without the need for manual intervention.
However, these general systems are not specifically designed for highly sparse recommendation models. 

To that end, Alibaba introduced XDL~\cite{jiang2019xdl}, an industry-scale training system designed for high-dimensional sparse data. XDL incorporates optimizations such as hierarchical sample compression, workflow pipelining, zero copy and CPU binding to improve training efficiency of the sparse part of the model. Kraken~\cite{sc20_kraken} targets at more efficient online training with decoupled key-value fetching and embedding, codesigned cache eviction policy with ML domain knowledge for the embedding tables, memory efficient optimizers for the sparse and dense part of the model, and a non-colocated deployment model allowing the inference servers and parameter servers to grow independently. ~\cite{sc20_intelDLRM} optimizes CPU-based DLRM training through lock-free embedding table update, tuned loop tiling for dense MLP, the \alltoall communication primitive and a new split-SGD implementation that takes advantage of the bits aliasing in FP32 and BFloat16 to reduce memory footprint. Baidu's AIBox~\cite{aibox} takes a different approach to horizontally scaling and focuses on fitting training of large recommendation models in a single node. AIBox hides serving latency by pipelining network, disk and CPU/GPU tasks, reduces model update overhead, and improves SSD life span through a grouped hashing scheme and a multi-level in-memory hashing system. 

Much attention is given to communication performance as it has become a major bottleneck in distributed training at cluster and datacenter scale. BytePS and ByteScheduler~\cite{byteps, bytescheduler} harnesses idle CPU and network resources and better communication scheduling to improve parameter exchange efficiency. However, in a homogeneous training cluster where each job spans multiple nodes, there are reduced opportunities for finding and exploiting spare network resources, resulting in a sub-optimal use of such approach. SwitchML and ATP~\cite{switchml, laoatp} leverages programmable network switches to perform in-network aggregation for cross-rack bandwidth reduction in datacenter environments.~\cite{luo2020plink, cai2021synthesizing} discovers and exploits datacenter network locality and forms optimized and dynamic aggregation routes through learning and optimal synthesis. Alternatively, these papers~\cite{lin2017deep, lim20183lc} address the communication overheads by using various quantization schemes to reduce communication volume.

\section{Conclusion}

DLRMs are an important class of models widely used by many internet companies for a wide range of applications. They can often be the single largest AI application in terms of infrastructure demand in data-centers. These models have atypical requirements compared to other types of deep learning models, but they still follow a similar trend of rapid rate of growth that is common across all deep learning-based applications. This growth constantly pushes the performance boundary required of the underlying software stack and hardware platform. 


In this paper we co-design a solution that  enables us to run models with trillions of parameters, while attaining $40\times$ faster total training time for production recommendation models. On the software side, \sys is equipped with a number of novel software techniques, including 4D parallelism, high-performance embedding kernels, hybrid kernel fusion, and hierarchical memory management. On the hardware side, the extensible \zionex platform allows for scaling up to the full data center with thousands of nodes, thus enabling a data center-scale AI training cluster to continue catering to the growing demands of deep learning models. 


Finally, we also explore co-designing models and algorithms to make them more amenable to the training cluster, for instance model architectures that reduce global \alltoall communication for better scaling efficiency. With this solution successfully deployed in production, we intend to continue working on these future directions to further push the capability for large scale deep learning training.

\section*{Acknowledgements}
We would like to acknowledge all of the help from members of the  hardware, datacenter and infrastructure teams, without which we could not have achieved any of the above reported results. This includes among others \emph{Jenny Yu, Matt Hoover, Hao Shen, Damien Chong, Jeff Puglis, Garnnet Thompson, Peter Bracewell, Anthony Chan, Wei Zhang, Michael Haken, Tiffany Jin, Joshua Held, Cheng Chen, Yin Hang, Ben Kim, Tyler Hart, Gada Badeer, Ahmed Qaid, Peichen Chang, Zhengyu Yang, Anil Agrawal, Viswesh Sankaran, Daniel Montgomery, James Taylor, Jeff Anderson, Amithash Prasad, Patrick Williams, Harsha Bojja, Arrow Luo, Changduk Kim, James Le, Rachel W Wang, Vignesh Mathimohan, Shockely Chen, Doug Wimer, James Allen, Vidya Rajasekaran, Kelly Zuckerman, Wenyin Fu, Valentin Andrei, Matt Skach, Philipp Keller, Olivier Raginel, Danielle Costantino.} We also like to thank other reviewers who have gone through multiple drafts of this paper, providing helpful inputs.


\bibliographystyle{ACM-Reference-Format}
\bibliography{refs}


\newpage
\onecolumn

\section*{Appendix-A}
\label{app:A}

\subsection*{Compute Benchmarks}

We collected and developed a set of operator-level benchmarks which we have also open sourced as part of \emph{PARAM bench}\footnote{\url{https://github.com/facebookresearch/param}}, 
to evaluate the representative problem sizes and shapes on the candidate hardware platforms and to better understand the throughput and latency in compute, memory, and communications.

\subsubsection*{GEMM benchmark} This benchmark calls cuBLAS GemmEx routine to compute matrix multiplications on configurable problem sizes with multiple precision choices. On the V100 GPU, this benchmark supports FP32 GEMM on the CUDA core and FP16 mixed-precision GEMM on Tensor Core. On the A100 GPU, it additionally supports TF32 GEMM and BF16 GEMM on the Tensor Core.

\begin{figure*}[h]
    \centering
    \includegraphics[width=\textwidth]{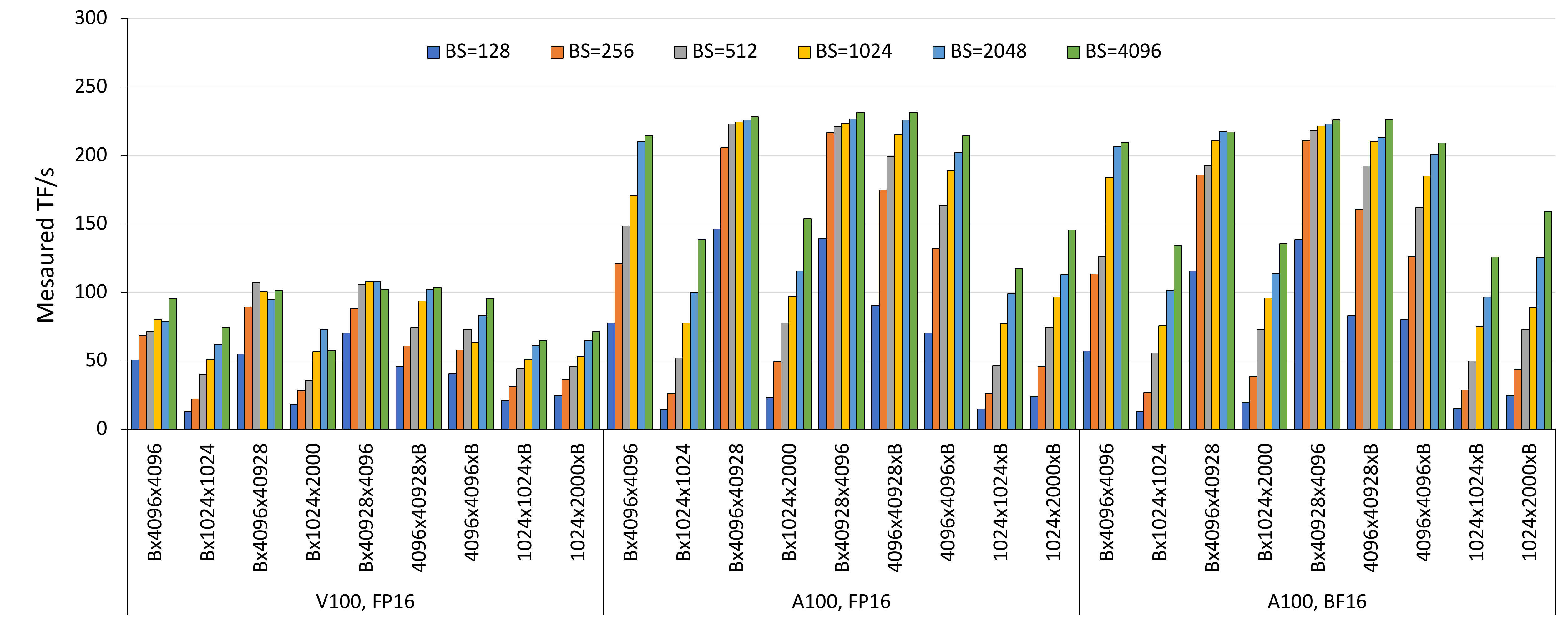}
    \caption{GEMM performance (TF/s) for V100 FP16 vs. A100 FP16/BF16.}
    \label{fig:gemm_16}
\end{figure*}

The benchmark results are shown in Figures~\ref{fig:gemm_16}.

\subsubsection*{MLP benchmark}

This benchmark implements the following multilayer perceptron (MLP) layers:
\begin{itemize}
\item Batch size = 128, 256, 512, 1024, 2048, 4096;
\item 20 MLP layers, where each layer is 1K$\times$1K , 2K$\times$2K and 4K$\times$4K;
\item Each layer has ReLU and final layers has SoftMax;
\item  Both backward and forward passes, including SGD update as the optimizer after the backward pass;
\item Precision support: FP16, BF16, TF32, FP32.
\end{itemize}

\begin{figure*}[h]
    \centering
    \includegraphics[width=\textwidth]{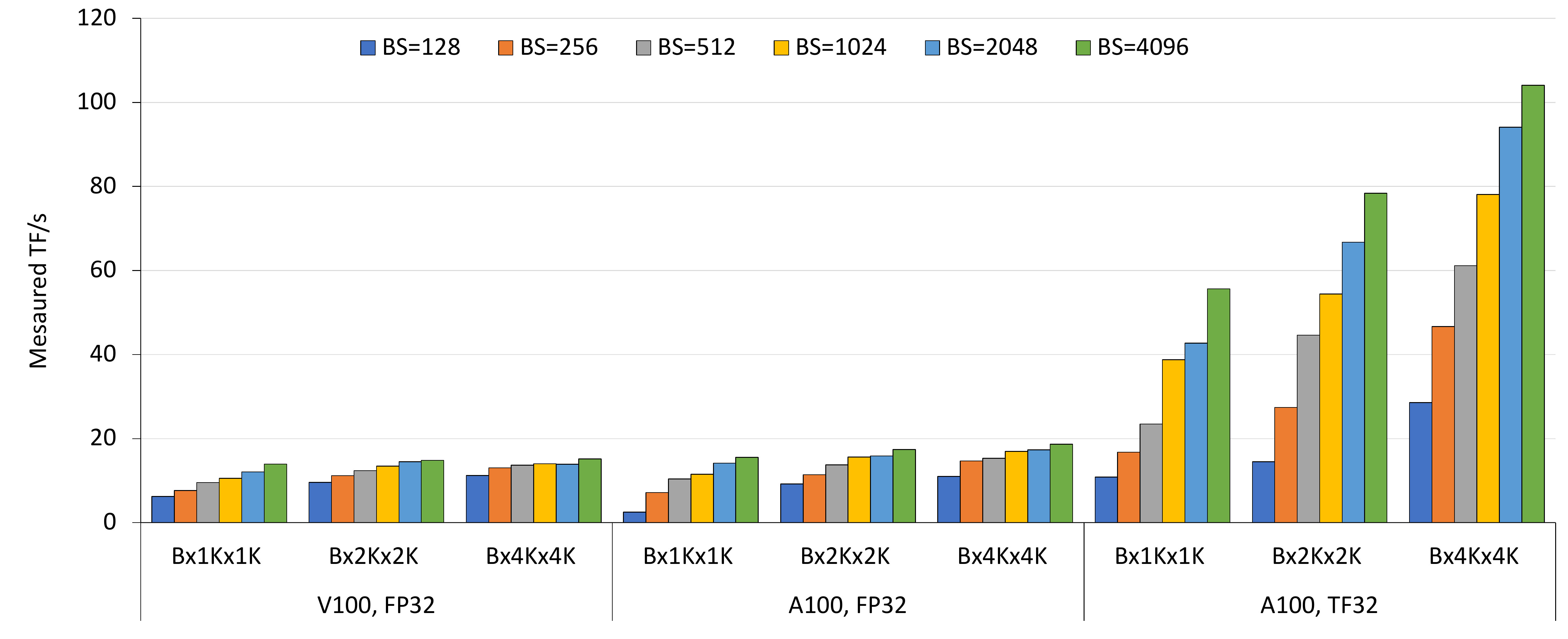}
    \caption{MLP performance for V100 FP32 vs. A100 FP32/TF32.}
    \label{fig:mlp_32}
\end{figure*}

\begin{figure*}[h]
    \centering
    \includegraphics[width=\textwidth]{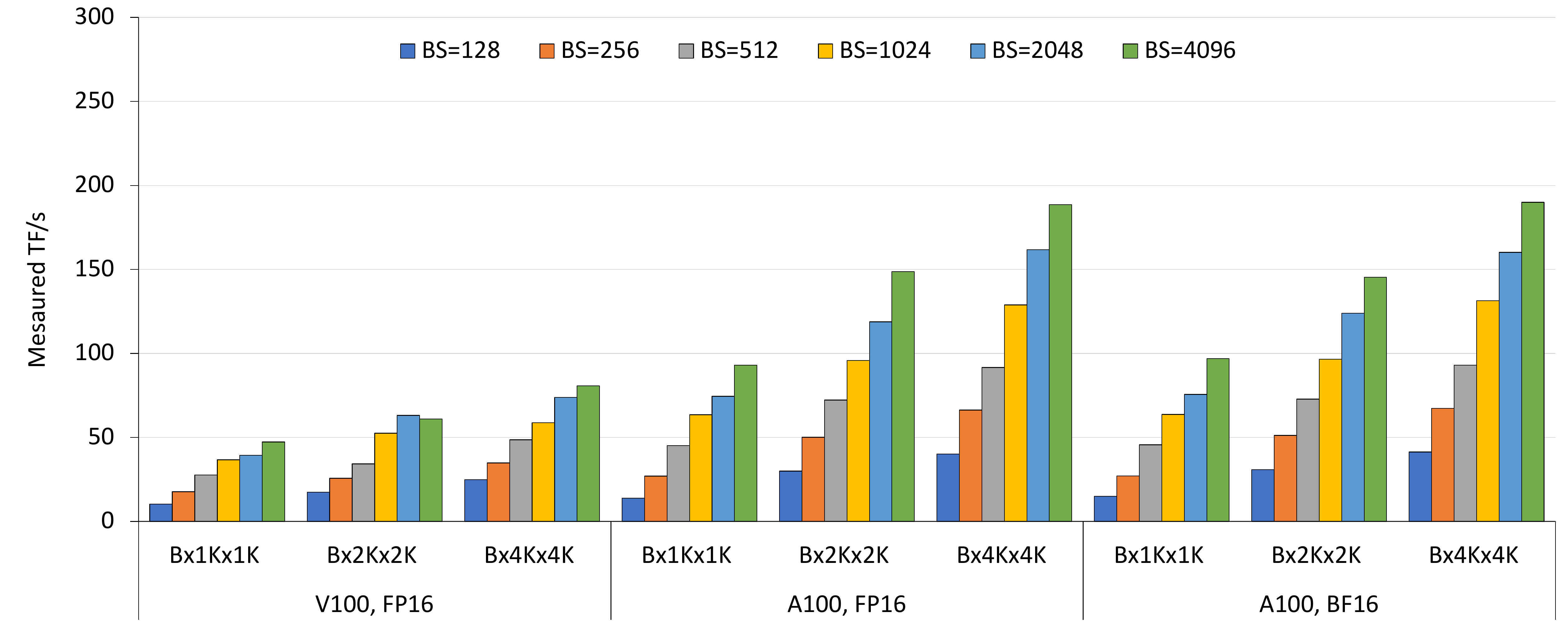}
    \caption{MLP performance for V100 FP16 vs. A100 FP16/BF16.}
    \label{fig:mlp_16}
\end{figure*}

The batch size, layer dimension, and number of layers can be configured to the customized number. We implemented this MLP benchmark using C++, directly implementing FC and FCGradients in the MLP layer using cuBLAS SGEMM/GemmEx function, ReLU with cuDNN cudnnActivationForward/ cudnnActivationBackward function, SoftMax with cudnnSoftmaxForward in the forward a customized CUDA kernel for the backward pass, and SGD optimizer with cuBLAS {\tt axpy} function. This benchmark can be used to project the performance of V100/A100 GPUs using a minimal MLP network without the framework overhead in PyTorch.
The benchmark results are shown in Figures~\ref{fig:mlp_32} and \ref{fig:mlp_16}.


\subsection*{Memory Benchmark}

This benchmark evaluates the achieved memory bandwidth of the embedding kernels described in Section~\ref{sec:embedding_ops}. To eliminate the L2 cache effects, a random tensor with 40 MB data (A100 L2 cache size) is allocated to flush the cache.

\begin{itemize}
\item Support the evaluation forward and backward pass (the backward pass is fused with optimizer);
\item Precision Support: FP32 and FP16;
\item Number of rows: 1000000, Number of tables: 64, Embedding dimension: 128, Pooling size: 32, rows per thread block: 32.
\end{itemize}

\begin{figure*}[h]
    \centering
    \includegraphics[width=\textwidth]{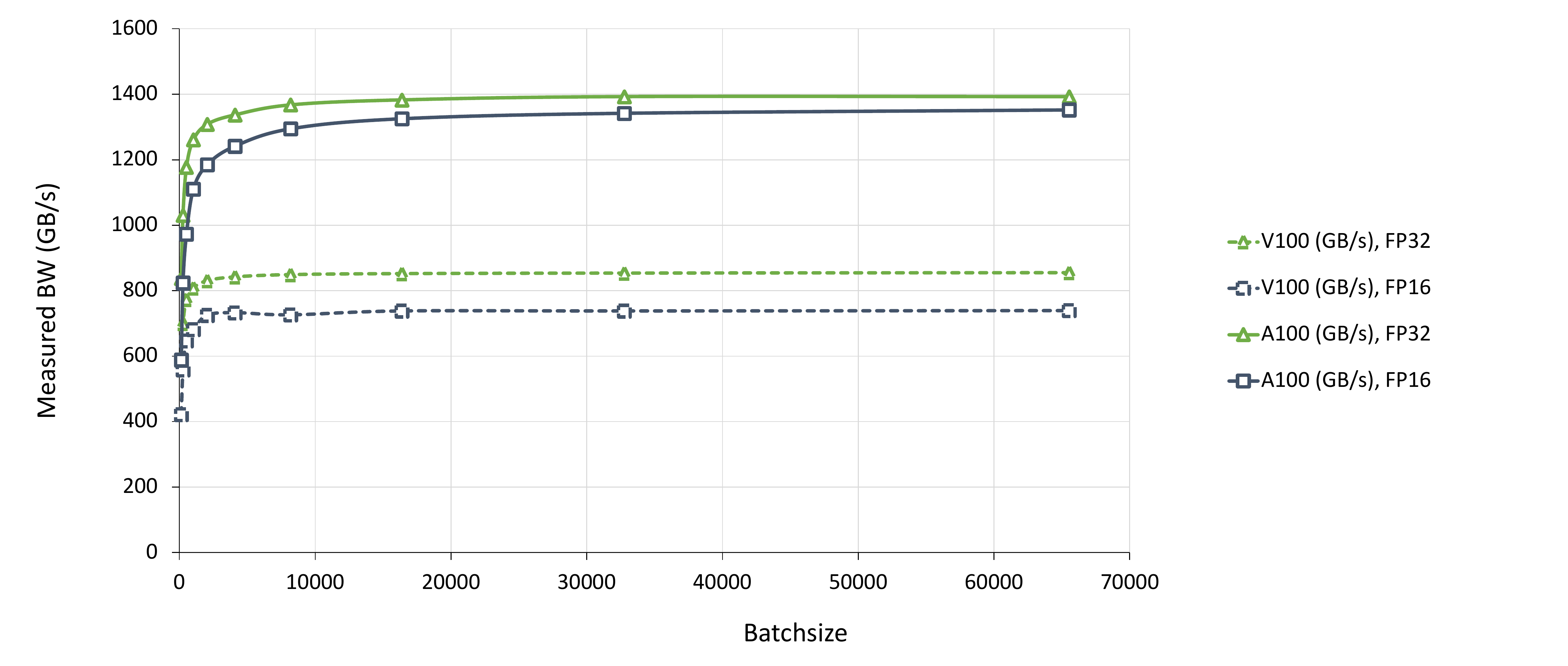}
    \caption{Achieved embedding lookup forward bandwidth using FP32 vs. FP16 on V100 vs. A100.}
    \label{fig:emb_fwd}
\end{figure*}

\begin{figure*}[h]
    \centering
    \includegraphics[width=\textwidth]{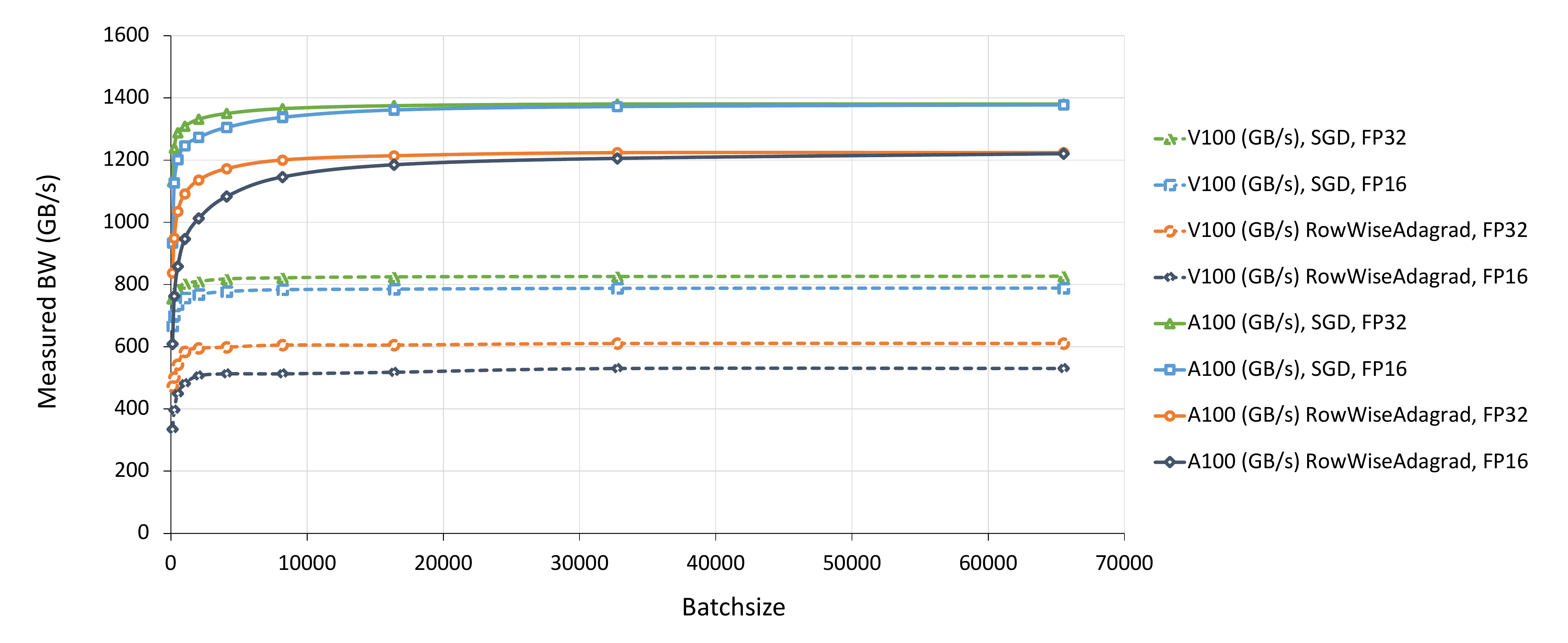}
    \caption{Achieved embedding lookup backward+optimizer bandwidth using FP32 vs. FP16 on V100 vs. A100.}
    \label{fig:emb_bwd}
\end{figure*}

The benchmark results are shown in Figures~\ref{fig:emb_fwd} and \ref{fig:emb_bwd}.

\subsection*{Communications Benchmark}

Low-level collective communication benchmarks, e.g. NVIDIA’s NCCL tests or OSU MPI benchmarks, have the following limitations:
\begin{itemize}
    \item Do not capture the behavior of actual workloads, i.e. exact message sizes, sequence of collective operations, etc. Instead these benchmarks support power-of-two message sizes - helpful to detect network trends.
    \item Limited to one specific communication library. As the name suggests, NCCL tests works only with NCCL and OSU MPI benchmarks is limited to MPI.
\end{itemize}

The PARAM comms benchmarks addresses these gaps by:
\begin{itemize}
    \item Creating common abstractions across platforms (e.g. NVIDIA GPUs, x86 CPUs, Google TPU etc.) to help standardize the benchmarking logic.
    \item Using PyTorch Process Group APIs to provide a portable interface across different communication libraries (e.g. NCCL, MPI, and UCC).
\end{itemize}

PARAM comms benchmarks supports two types of collective benchmarks:
\begin{itemize}
    \item Bench mode: Simplest mode of operation similar to NCCL tests. Run single collective in blocking or non-blocking manner across fixed set of message sizes (e.g. power of 2 message sizes). This is mainly used for low-level HW testing
    \item Replay mode: Replays a trace of collective communication calls to mimic exact workload behavior in terms of collective sizes. 
\end{itemize}

\begin{figure*}[h]
    \centering
    \includegraphics[width=\textwidth]{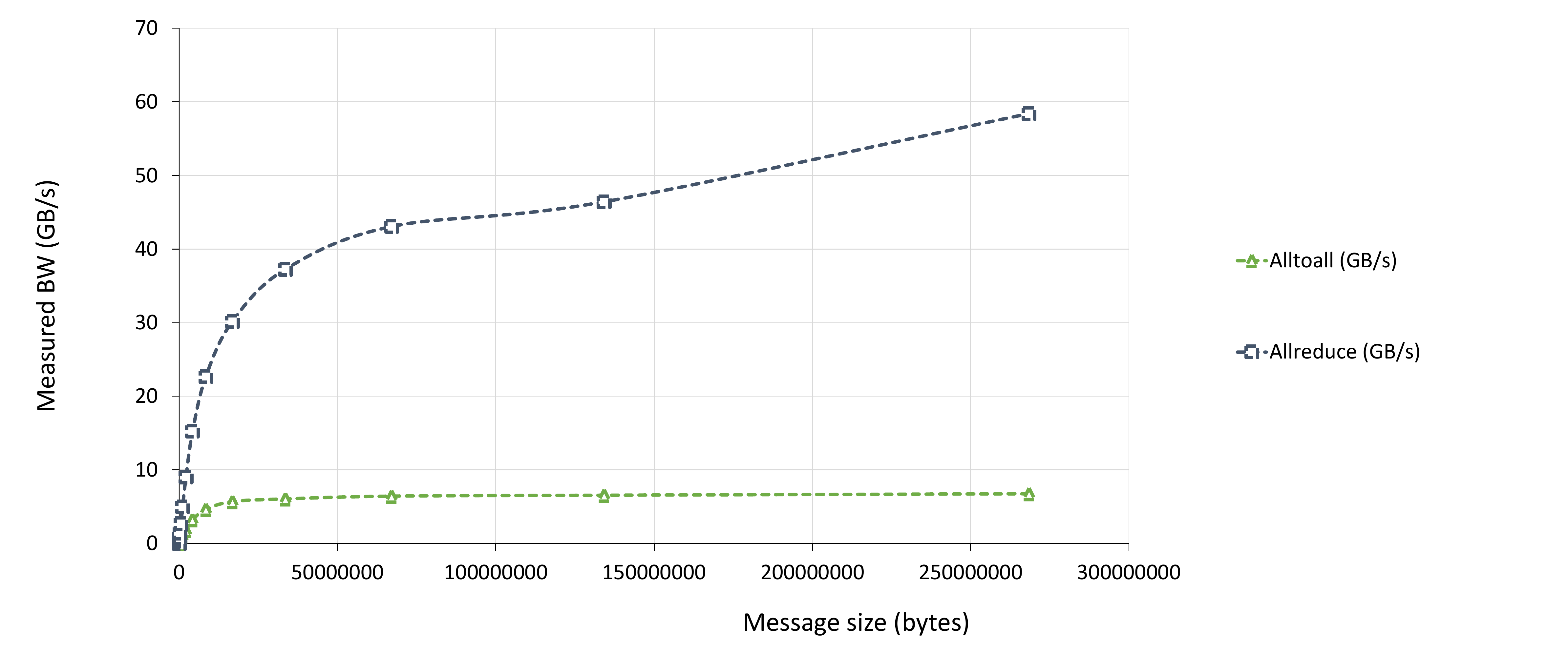}
    \caption{Achieved \alltoall and \allreduce bandwidth at 128GPUs}
    \label{fig:comms}
\end{figure*}

Figure \ref{fig:comms} presents \alltoall and \allreduce benchmark scaling for power-of-two message sizes on 128 GPUs. \alltoall achieves ~7GB/s and is primarily limited by scale-out bandwidth (12.5 GB/s peak; 10.5 GB/s achievable on V100). \allreduce achieves higher bandwidth since it uses NVLINK more effectively. 
\\

\newpage
\section*{Appendix-B}
\label{app:B}

\subsection*{Performance roofline and benchmarking}
\label{sec:perf_model}

\begin{wrapfigure}{r}{0.45\textwidth}
    \vspace{-5mm}
    \centering
    \includegraphics[width=0.25\textwidth]{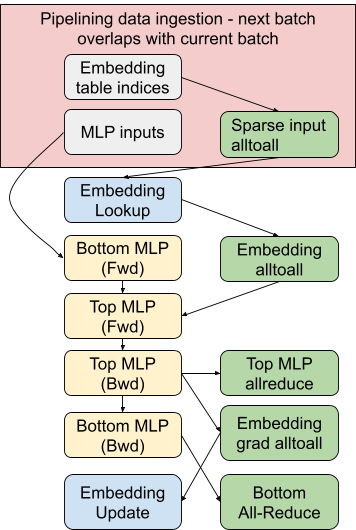}
    \caption{DLRM dependency graph}
    \label{fig:perfModel}
\end{wrapfigure}

In order to identify performance gaps, to see how far we are from fully utilizing the platform capabilities - we establish the upper bound for achievable performance using an analytical roofline model. DLRMs can be broken down into the following major components - 1) bottom MLP; 2) embedding lookup and update; 3) \alltoall communication of the model-parallel pooled embeddings; 4) interaction and Top MLP; 5) \allreduce communication for the data-parallel MLP gradient synchronization. The execution dependency between there different components are outlined in Fig.\ref{fig:perfModel}. As discussed above, individually each of these have different characteristics. The latency/performance for each component is dependent on different parts of the system, for instance the embedding ops performance depends on the achievable HBM bandwidth, whereas the MLP performance is bounded by achievable compute flops. Even between the two collective communication primitives - \allreduce performance  depends on both the scale-out and scale-up bandwidths, whereas the \alltoall performance primarily depends on the scale-out bandwidth. With estimates for latencies for these individual components, the overall per-iteration latency can be estimated as shown in Eq.~\ref{eq:train_latency}

\begin{gather*}
T_{fwd} = \max[BotMLP_{fwd}, (Embedding\_lookup + alltoall_{fwd})] + Interaction_{fwd} + TopMLP_{fwd} \\
T_{bwd} = \max[TopMLP_{bwd} + Interaction_{bwd} + \max\{alltoall_{bwd} + Embedding\_update, BotMLP_{bwd}\}, \\ (TopMLP\_Allreduce + BotMLP\_Allreduce)]
\end{gather*}
\begin{gather}
\label{eq:train_latency}
T_{total} = T_{fwd} + T{bwd}
\end{gather}

To estimate the performance and latencies for each of these components, we use operator-level benchmarks which allow evaluation of target operator shapes/sizes on candidate HW platforms. We benchmark\footnote{\url{https://github.com/facebookresearch/param}} 
the 1) embedding operators, 2) typical MLP sizes, and 3) communication primitives. With these benchmarks we are able to establish the max achievable HBM bandwidth to be 850 GB/s for V100 and 1300 GB/s on A100 GPUs, and for the MLP sizes of interest, achievable compute efficiencies to be up to 78.6\% (V100) and 70.5\%. (A100). Furthermore, we achieve 7GB/s for 256MB \alltoall and 60GB/s for 256MB \allreduce. \allreduce is able to achieve higher effective bandwidth since it utilizes both scale-out and NVLINK badnwidths. These benchmarking results and configuration used are detailed in \href{app:A}{Appendix-A}. 

\end{document}